\documentclass[a4paper,11pt]{article}
\pdfoutput=1 

\usepackage{jheppub} 
\usepackage[mathscr]{euscript}
\usepackage[T1]{fontenc} 

 \usepackage{musicography}
 
 \newcommand{\sO}{\mathscr {O}}
 
\usepackage{bm}
\usepackage{verbatim}
 
 \usepackage{makecell}
 \usepackage{float}
 
 \usepackage[makeroom]{cancel}
 
 \newcommand{\bv}{ \begin{verbatim}}

\newcommand{\be}{\begin{equation}}
\newcommand{\ee}{\end{equation}}
\newcommand{\bpm}{\begin{pmatrix}}
\newcommand{\epm}{\end{pmatrix}}

\newcommand{\EV}[1]{\langle #1 \rangle}
\newcommand{\beqn}{\begin{eqnarray}}
\newcommand{\eeqn}{\end{eqnarray}}

\usepackage{slashed}

\newcommand{\cG}{\mathcal G}

\newcommand{\cO}{\mathcal O}

\newcommand{\cN}{\mathcal N}
\newcommand{\cM}{\mathcal M}

\newcommand{\sfs}{{\sf s}} 
\newcommand{\sft}{{\sf t}} 
\newcommand{\sfu}{{\sf u}}

\newcommand{\sfG}{\mathsf G}

\newcommand{\dDisc}{{\text{dDisc}}}

\newcommand{\sfC}{{\sf C}}
\newcommand{\sfJ}{{\sf J}}

\DeclareMathOperator{\Real}{Re}
\DeclareMathOperator{\Res}{Res}

\author[\natural]{Hongliang Jiang }

 \affiliation[\natural{}]{Centre for Theoretical Physics, School of Physical and Chemical Sciences, \\
Queen Mary University of London,\\
 Mile End Road, E1 4NS,  UK}

 \emailAdd{h.jiang@qmul.ac.uk}

\preprint{QMUL-PH-22-23}

\title{\boldmath\huge Celestial Mellin Amplitude}

\abstract{Celestial holography provides a  promising avenue to studying bulk   scattering in flat spacetime from the perspective of boundary celestial conformal field theory (CCFT).  A key ingredient in connecting the two sides is the celestial   amplitude, which is given by the Mellin transform of momentum space scattering amplitude in energy. As such,  celestial   amplitudes can be identified with the correlation functions in celestial conformal field theory. 
In this paper, we introduce the further  notion of celestial Mellin amplitude, which is given  by the   Mellin transform of  celestial   amplitude in coordinate. 
For technical reasons, we focus on the celestial Mellin amplitudes for scalar fields in  three dimensional flat spacetime dual to 1D CCFT, and discuss the  celestial Mellin   block expansion. In particular, the poles of the celestial Mellin amplitude encode  the scaling dimensions of the possible exchanged operators, while the residues there are related to the OPE coefficient squares in a linear and explicit way. 
We also compare the celestial Mellin amplitudes with the coefficient functions which can be  obtained using inversion formulae. 
Finally, we make some comments about the possible  generalizations of   celestial Mellin amplitudes to higher dimensions.
}

\begin{document} 
\maketitle
\flushbottom
\allowdisplaybreaks

\section{Introduction}

Over the past few years, there are some dramatic progress in studying holography for quantum gravity in the absence of cosmological constant. Such a new program of holographic duality is dubbed  celestial holography. The crucial point of celestial holography stems from the fact that the Lorentz group in $d$-dimensional Minkowski spacetime  $SO(d-1,1)$ can be identified with the conformal group in $(d-2)$ dimensions.  Moreover, near the null boundary of flat spacetime, the symmetry group is even  enhanced to the infinite dimensional asymptotic symmetry, called Bondi–van der Burg–Metzner–Sacs (BMS) symmetry~\cite{Bondi:1962px,Sachs:1962wk}. 
As shown in  \cite{Strominger:2013jfa,He:2014laa},  the infinite  dimensional BMS symmetry can be used to understand the  soft theorems of graviton scattering amplitudes.  Since then, many interesting results related to the infrared structure of gravity were discovered,  see \cite{Strominger:2017zoo} for a review. Furthermore, more and more non-trivial evidence  were accumulated, which indicate  that   the  $d$-dimensional   quantum gravity in flat spacetime can be described in terms  of $(d-2)$-dimensional conformal field theory, called celestial conformal field theory (CCFT)  living at null infinity. 
Just like the well-known AdS/CFT correspondence, such a  kind of relation between bulk gravity and boundary field theory  is   reminiscent of the principle of holography, and is referred to as celestial holography.  
The connection between the bulk and boundary can be understood more directly by introducing the  so-called celestial amplitude~\cite{Pasterski:2016qvg,Pasterski:2017kqt}, which is nothing but the Mellin transform  of the momentum space scattering amplitude. Consequently, one can regard the celestial amplitudes as the correlation functions in CCFT. 
Nowadays,   celestial holography has become a very rich and interesting subject. See   \cite{Pasterski:2021rjz,Raclariu:2021zjz}  for a review. 

As such,   understanding   quantum gravity in flat Minkowski spacetime boils down to the study of celestial conformal field theory. Compared to the standard CFT,   CCFT shares  lots of similarities but also features many peculiarities. Little is known about the fundamental axioms for   defining   CCFT, and most of the understanding of CCFT is based on symmetry.  For example, the notion of unitarity in CCFT is not clear. This should inherit from the unitarity of bulk quantum gravity, and is very different from  unitarity in conventional  CFTs. As a result, the scaling dimension of operators in CCFT can be even complex, whose implication is still vague. Nevertheless, 
one may hope to  borrow some techniques  from standard CFT, and try to understand some aspects of CCFT.  
A notable success is the operator product expansion (OPE) in CCFT, which arises from the collinear limit of bulk scattering amplitudes \cite{Fan:2019emx,Pate:2019lpp,Jiang:2021csc,Adamo:2021zpw}. This enables us to understand the short distance behaviors and operator algebras in CCFT. Moreover, it also helps us to reveal  the  hidden new structure of CCFT, like  the $w_{1+\infty}$ symmetry \cite{Guevara:2021abz,Strominger:2021lvk, Himwich:2021dau,Jiang:2021ovh} which is difficult to anticipate using standard techniques. 

A natural question is how far can we go by importing the techniques from standard CFTs? 
In particular, recently there is a huge amount of progress  in studying standard CFTs both analytically and numerically, under the name of conformal bootstrap. Many interesting and powerful techniques are developed.  One of them is the Mellin amplitude \cite{Mack:2009mi}. 
The Mellin  techniques are very useful in   bootstrapping CFTs both perturbatively and non-perturbatively.
  The Mellin amplitude also offers an excellent tool in studying holographic CFTs and thus the corresponding AdS  quantum gravity \cite{Fitzpatrick:2011ia}. It can be regarded as the natural AdS analogue of flat space scattering amplitude, and the latter can be   recovered from Mellin amplitude in a suitable limit \cite{Penedones:2010ue}. The Mellin amplitude has simple  poles corresponding to the exchanged operators, and 
 the  crossing symmetry of the CFT correlator maps to the amplitude crossing symmetry in Mellin space.  A very successful application of Mellin techniques is the calculation of holographic correlators in $AdS_5\times S^5$, which is   simple when written in the Mellin basis, and can be bootstrapped based on general consistency conditions and maximal supersymmetry  without worrying  about the complicated interactions of bulk supergravity \cite{Rastelli:2017udc}. 

Considering the powerful role of Mellin space in AdS/CFT, in this paper, we would like to   import the techniques of Mellin amplitudes to celestial holography. More precisely, we will introduce the notion of celestial Mellin amplitude.  
Recall that starting with  momentum space scattering amplitudes, one can perform the Mellin transformations in energies and obtain the so-called   celestial amplitudes. We will  also refer to celestial amplitudes as  celestial correlators, because they are indeed identified with the correlators in CCFT.  As in standard  CFT, we can further consider the  
  Mellin amplitudes for the celestial correlators in CCFT,
 and this naturally leads to the   definition of celestial Mellin amplitudes. The relationship of various kinds of amplitudes in celestial holography is then   illustrated below:
\begin{figure}[h]
\includegraphics[width=1.\textwidth ]{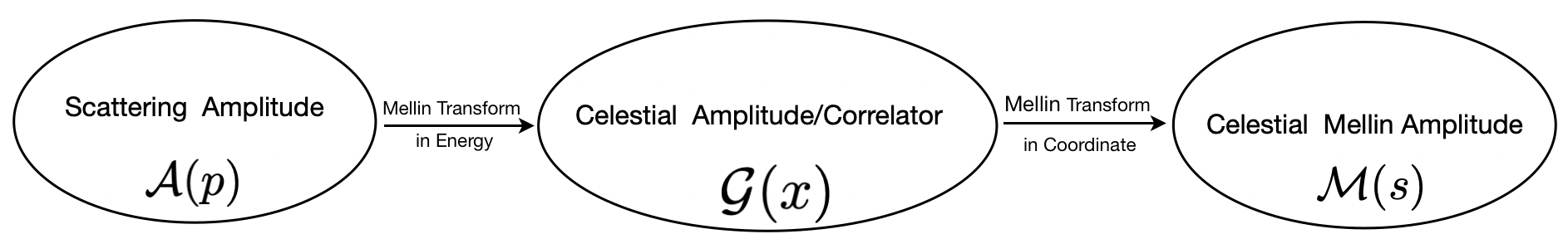}
\end{figure}

Although the Mellin amplitudes have been well studied and understood in standard CFTs, the implementation of Mellin amplitudes in CCFTs suffers   various problems. A notable feature of    CCFT correlators is that they come  with some kinematical constraints arising from momentum conservation. For example, in four and higher dimensional bulk spacetime, the four-point correlators    always contain a delta function, which enforces the positions of  four operators to lie  on the intersection of the celestial sphere and the scattering plane.    This delta function   makes the analytic structures   of correlation functions obscure and  is absent in standard CFTs.  Therefore it is not obvious to what extent it is meaningful to directly apply the techniques of Mellin amplitude in standard CFT. 
Instead, here we will be mainly focusing on the three dimensional spacetime where the CCFT lives on the one-dimensional celestial circle.  The restriction to  three dimensions then avoids the  problem of kinematical constraint,  and   the delta function is absent in  the   four-point celestial correlators of 1D CCFT.  As a result, it seems that we can then apply the standard Mellin techniques to our one dimensional CCFT.  However, the standard Mellin amplitude   is normally used for  CFTs in high  enough dimensions, and  seems to suffer  from some  subtleties in one dimension due to additional kinematical constraints. \footnote{Nevertheless, \cite{Ferrero:2019luz} implemented such a kind of Mellin technique  in  bootstrapping 1D CFT.  } To further avoid these subtleties, we will use the Mellin techniques   specially designed for 1D CFT in~\cite{Bianchi:2021piu}.  In such a case, the Mellin amplitude is literally given by the Mellin transform  in coordinate of  1D CFT correlators. This is the strategy that we will adopt in this paper. 

To be even more precise and for simplicity, in this paper, we will be mainly focusing on the celestial Mellin amplitudes corresponding to the scattering of  four identical scalar particles    in three dimensional spacetime. The celestial Mellin amplitude is given by the double Mellin transformations of momentum space scattering amplitude, one in energy \eqref{Gz4} and the other in coordinate \eqref{MellinTsf}. 
 We will also consider the celestial  Mellin block expansion, namely the decomposition of  celestial Mellin amplitudes into  the sum of     Mellin blocks for different exchanged operators.   A very nice feature of the celestial Mellin blocks is that they are by construction crossing symmetric. 
It turns out that the celestial Mellin amplitudes enable us to perform such an expansion easily.  In particular, the poles of celestial Mellin amplitudes encode  the scaling dimensions of possible exchange operators, while the residues there are related to the OPE coefficients in a specific way. We derive the   explicit formula \eqref{OPEcoeff} to compute the OPE coefficients directly from the residues of celestial Mellin amplitude  at various poles.  The resulting  celestial Mellin block expansion agrees with the conformal block expansion for 1D CCFT.  
We will also compare the celestial Mellin amplitudes with the coefficient functions which can be derived from Euclidean or Lorentzian inversion   formulae \cite{Caron-Huot:2017vep,Simmons-Duffin:2017nub}. \footnote{Lorentzian inversion formulae is another powerful CFT technique and also plays an important role in studying holographic CFTs. It would be very interesting to investigate its application in celestial holography.  } They   are related by integral transformation \eqref{CMAvsOPEcoff}.     The poles of the coefficient functions give precisely the scaling dimensions of exchange operators, while the residues there are precisely the OPE coefficient squares. However, coefficient functions are usually quite complicated and difficult to compute explicitly. This is in sharp contrast with celestial Mellin amplitudes which are much easier to calculate. 
We will demonstrate all these features  in a    very concrete example of scalar fields in 3D. 

Although we are mainly focusing on the celestial Mellin amplitude in 3D, we will also explore several possible generalizations to higher dimensions. As we discussed before, the main subtlety comes from the delta function constraint in the  four-point  correlators of CCFT.  In the simplest scenario, one may ignore this problem and blindly apply the Mellin   techniques in standard CFT. A better way to avoid the delta function is to perform the shadow or light transform on the CCFT correlators, and then study their corresponding Mellin amplitudes. The benefit of performing shadow or light transforms enables us to remove the delta functions, thus making the CCFT correlator better behaved.  
Finally, we may regard the four point correlators in CCFT as one dimensional  defect correlators, although the defect is actually trivial whose only role is   restricting the positions of operators. On such a one dimensional defect CFT, we can then   apply the    Mellin transform directly, just like the case of 3D bulk dual to 1D CCFT. All these different approaches have their pros and cons, and they are supposed to be related in a non-trivial way. 
We will not study their connections in detail here, but leave this important question to the future.

The rest of the paper is organized as follows. 
 In section~\ref{CMAin3D},  we   introduce the notion of celestial Mellin amplitude in 3D and discuss its block expansion. 
In section~\ref{CMAexample}, we  consider a very explicit example  of scalar fields in 3D, and compute its celestial Mellin amplitude as well as block expansion. We will also calculate the coefficient functions in order to compare with the celestial Mellin amplitude.
In section~\ref{CMAinhighD}, we   briefly comment on the generalization of celestial Mellin amplitude to higher dimensional spacetime. 
In section~\ref{conclusion}, we conclude and  discuss some possible future directions. 
In appendix~\ref{MNinverse}, we   prove a claim about the inverse of an infinite dimensional matrix, relating residues and OPE coefficients. 
 In appendix~\ref{AppInversion}, we include many technical details of the coefficients functions derived from     Euclidean and Lorentzian  inversion formulae.

\section{Celestial Mellin amplitude  in 3D} \label{CMAin3D}

 In this section, we introduce the notion of celestial Mellin amplitudes in 3D. We will first review the celestial amplitudes, and then define the  celestial Mellin amplitudes by performing another Mellin transformation in coordinate. We will also discuss the Mellin block expansions of celestial Mellin amplitudes, and  derive the key formula \eqref{OPEcoef} relating the OPE coefficients and residues of celestial Mellin amplitudes.  We will  also relate the celestial Mellin amplitudes with  the coefficient functions computed by inversion formulae. Finally, we will describe the Regge behavior of celestial Mellin amplitudes.

  \subsection{Celestial   amplitude}
Let us first introduce the celestial amplitudes. For our purpose, we will be mainly considering the massless scalar fields in (2+1)D. The starting point originates from the fact that 
 the Lorentz group in (2+1)D coincides with the conformal group in 1D, namely     $SO(2,1)\simeq SL(2,\mathbb R)$. To make the relation manifest, we can parameterize the momentum of each particle  as follows:   \footnote{We use the  signature  $(-++)$.} 
  \be
  p^\mu =\epsilon \omega q^\mu(   x)~ , \qquad  q^\mu=(1+x^2, 2x, 1-x^2)~, \qquad q \cdot q=0~.
  \ee
  where $\omega >0$ is the energy along null direction $q^\mu$, and $\epsilon $ labels whether the particle is   incoming ($\epsilon=-1$) or outgoing ($\epsilon=+1$).
Here $x\in \mathbb R$ is just the (stereographic) coordinate on the celestial circle. 
 It is easy to check that performing a global conformal transformation $SL(2, \mathbb R)$ on $x$ together with a corresponding transformation on $\omega$  leads to a bulk Lorentz transformation  $SO(2,1)$ on $p^\mu$. 
  
  In general, the scattering amplitude of $n$ massless scalar fields in (2+1)D can be written as
  \be\label{Amp}
  \mathcal A  (p_i) =A (p_i) \delta^3\Big(\sum_{i=1}^n  p_i
  \Big)~.
  \ee 
The delta function in \eqref{Amp} enforces the momentum conservation,   making the translational symmetry manifest. 

Now, we would like to have manifest Lorentz symmetry. This can be achieved by introducing   the so-called celestial amplitude, which is given by   Mellin transforming the  momentum space scattering amplitude  \cite{Pasterski:2016qvg,Pasterski:2017kqt}
  \be\label{CAmp}
  \cG( \Delta_i, x_i)  =\prod_{i=1}^n \int_0^\infty d\omega_i \; \omega_i^{\Delta_i-1} 
   \mathcal A\Big(p_j=\epsilon_j \omega_j q^\mu(x_j)\Big) ~.
  \ee
   
  The nice property of celestial amplitude is that it makes the Lorentz symmetry manifest. More specifically,   the Lorentz transformation in the (2+1)D bulk induces a conformal transformation on the celestial circle, under which the celestial amplitude transforms as
  \be
      \cG\Big( \Delta_i, \frac{a x_i+b}{c x_i+d}\Big)
      =\prod_j  (  {c x_j+d}  )^{2\Delta_j}  \;  \cG( \Delta_i, x_i)  ~, \qquad ad-bc=1~,
  \ee
 which is nothing but the transformation rule of   CFT correlators. Therefore  the celestial amplitudes can be regarded as the correlation functions of some operators in    CCFT 
  \be
    \cG( \Delta_i, x_i) =\EV{\cO^{\epsilon_1}_{\Delta_1}(x_1)\cdots \cO^{\epsilon_4}_{\Delta_4}(x_4)}_{\text{CCFT}}~.
  \ee
  In particular, $\Delta_i$ is just the scaling dimension of operator    in CCFT.

Let us now specialize the discussion to the case of four particle scattering. 
 In this case, it is convenient to define  the Mandelstam variables 
   \be
  \sfs=-(p_1+p_2)^2~, \qquad 
 \sft=-(p_1+p_3)^2~, \qquad   \sfu=-(p_1+p_3)^2 ~,
  \ee
  which are subject to the condition $\sfs+\sft+\sfu=0$ since all the external particles are massless. 
  Then the momentum space amplitude can be written in terms of Mandelstam variables 
  \be \label{Amp4}
  \mathcal A (p_i) =A (\sfs,\sft) \delta^3\Big(\sum_{i=1}^4  p_i   \Big)~.
  \ee   
   In order to compute the celestial amplitude, we would like to simplify \eqref{CAmp}. 
    The delta function of  momentum conservation in \eqref{Amp4} enables us to perform the three among four   integrals in \eqref{CAmp}, enforcing the constraint
  \be
  \frac{\omega_2}{\omega_1}=- \frac{\epsilon_2}{\epsilon_1}\frac{x_{13}x_{14}}{x_{23}x_{24}}~, \qquad
    \frac{\omega_3}{\omega_1}=  \frac{\epsilon_3}{\epsilon_1}\frac{x_{12}x_{14}}{x_{23}x_{34}}~, \qquad
      \frac{\omega_4}{\omega_1}=- \frac{\epsilon_4}{\epsilon_1}\frac{x_{12}x_{13}}{x_{24}x_{34}}~.  \qquad
  \ee
  where $x_{ij}=x_i-x_j$. As a result, we can express
  \be
 \sfs=-4x_{12}^2\frac{x_{13}x_{14}}{x_{23}x_{24}} \omega_1^2~, \qquad
  \sft=-\frac 1 z \sfs~, \qquad
   \sfu=\frac{1-z}{z}\sfs~,
  \ee
   where we introduce the cross-ratio
  \be
  z=\frac{x_{12}x_{34}}{x_{13}x_{24}}\in \mathbb R~.
  \ee
  For simplicity, let us further assume that all   the conformal dimensions are the same, $\Delta_i=\Delta_\phi$. Then the celestial amplitude has the following structure \cite{Lam:2017ofc}
 \be
    \cG(  x_i)  =\frac{1}{|x_{12}|^{2\Delta_\phi}|x_{34}|^{2\Delta_\phi}}
    G(z)~,
  \ee
where
\be\label{Gz4}
G(z) = 2^{-4\Delta_\phi+1}\frac{|z|}{\sqrt{|z-1|}}
\int_0^\infty d\omega\; \omega^{4\Delta_\phi-4}A\Big( \epsilon_1 \epsilon_2 \omega^2, -\frac1z\epsilon_1 \epsilon_2 \omega^2\Big)~.
\ee
One important feature of four-point celestial amplitude in (2+1)D is the absence of delta function, which is inevitable in higher dimensions due to kinematic constraints. 

In the four particle scattering, there are various channels depending on which particle is incoming or outgoing. In different channels, the function $G(z)$ takes different forms, so we can write
  \be
G(z)=\begin{cases}
G^{(-)}(z)~, & z\in (-\infty, 0)~, \qquad  14\leftrightarrow 23 \text{ channel}~, \\
G^{(0)}(z)~, & z\in ( 0,1)~, \qquad\quad\;  13\leftrightarrow 24 \text{ channel}~, \\
G^{(+)}(z)~, & z\in (1,\infty )~, \qquad\quad\!  12\leftrightarrow 34 \text{ channel} ~,\\
\end{cases}
  \ee
where  we use arrows to distinguish incoming and outgoing particles; for example $12\rightarrow 34$ indicates that $1,2$ are incoming, while $3,4$ are outgoing, thus in this case $\epsilon_1=\epsilon_2=-\epsilon_3=-\epsilon_4=-1$. 

For the scattering of four identical particles, we have crossing symmetry  which relates scattering amplitudes in different channels:
\be\label{Astu}
A(\sfs,\sft)=A(\sft,\sfs) =A(\sfu,\sft)~, \qquad \sfs+\sft+\sfu=0~.
\ee
Using eq.~\eqref{Gz4}, one can show that the crossing equations above lead  to the following identities  \cite{Lam:2017ofc}
\beqn\label{CFTcrossing}
G^{(-)}(z) &=&G^{(0)}\Big(\frac{z}{z-1}\Big)~, \qquad  z\in (-\infty,0) ~,\\
G^{(+)}(z) &=& z^{2\Delta_\phi}G^{(0)}\Big(\frac{1}{z}\Big)~, \qquad  z\in (1,\infty) ~,\label{CFTcrossing2} \\
G^{(0)}(z) &=& \Big(\frac{z}{1-z}\Big)^{2\Delta_\phi}G^{(0)}(1-z)~,  \qquad  z\in (0,1)   ~.
\label{CFTcrossing3}
\eeqn
These are nothing but the equations arising from the Bose symmetry in 1D CFT by exchanging $ 1\leftrightarrow 2, 2\leftrightarrow 3, 2\leftrightarrow 4$, respectively \cite{Mazac:2018qmi}.  These equations also allow us to find the full profile of $G(z)$ once we know $G^{(0)}(z)$ in the range of $0<z<1$.
 
   
 \subsection{Celestial Mellin amplitude}
 In this subsection, we  introduce the celestial Mellin amplitude by performing a further Mellin transform  in coordinate. 
 
 As we discussed before, the most interesting feature of celestial amplitudes is that they can be regarded as the correlators in CCFT.  In the special case of four point celestial amplitude \eqref{Gz4}, we can thus identify it as the correlation function of four identical scalar operators with dimension $\Delta_\phi$ in CCFT: \footnote{Rigorously speaking, the four scalar operators are not identical, because some are incoming, some are outgoing. However, this turns out to be irrelevant in the discussion below, so we will ignore this minor difference. }
 
\be\label{cGCFT}
\cG(x_i)=\EV{\phi(x_1) \cdots \phi(x_4)}_{\text{CCFT}}=\frac{1}{|x_{12}|^{2\Delta_\phi}}\frac{1}{|x_{34}|^{2\Delta_\phi}}G(z)~.
\ee
The function $G(z)$ satisfies \eqref{CFTcrossing}-\eqref{CFTcrossing3}. Note that non-trivial 1D CFT is always non-local.~\footnote{1D CFT can arise from higher dimensional CFTs by restricting to the 1D defect, namely setting $z=\bar z$,  or appear  on the boundary of $AdS_2$. }
Nevertheless, in the conventional situation, the 1D CFT is assumed to be unitarity. 
For CCFT arising from celestial holography, it is not unitary anymore in the standard sense. For example, the scaling dimension of operators in CCFT can be  an arbitrary complex number.~\footnote{However, it is worth emphasizing that there should a notion of unitarity induced from the unitarity in the bulk quantum fields. A better understanding of such a kind of unitarity is crucial in celestial holography.}

 In the following discussion, we will be mostly focusing on the $13\leftrightarrow 24$ channel with  $0<z<1$. Starting with $G^{(0)}(z)$ in the range  $0<z<1$,  one can  analytically continue the function $G^{(0)}(z)$ to  the    complex value. Then one can show that, at least for the standard   1D unitary CFT,  the function $G^{(0)}(z)$ is holomorphic on  $\mathbb C\backslash\{ (-\infty,0)\cup (1, \infty) \}$, where  $(-\infty,0)$ and $ (1, \infty)$ are the branch cuts.
   For simplicity of notation, we will just write $G^{(0)}(z)$ as $G(z)$ below, unless it causes confusion. 

Then we have the crossing equation \eqref{CFTcrossing3},  arising from the symmetry of    exchanging of $x_1$ and $x_3$:
\be\label{crossingeq}
 z^{-2\Delta_\phi}G (z )
 = (1-z)^{-2\Delta_\phi}G (1-z )~,
   \qquad  z \in (0,1 ) ~.
\ee 

Now we would like to discuss the conformal block expansions of four-point CFT correlators.    Recall that for 1D CFT, the   $sl(2, \mathbb R)$ global conformal block reads 
\be\label{confblocks}
G_\Delta(z) =z^\Delta {}_2F_1(\Delta, \Delta; 2\Delta; z)~.
\ee

Then in the limit $z\to 0$, namely $x_1 \to x_2$,  we can consider the $s$-channel conformal block expansion
\be\label{confdecomp}
G(z) =\sum_{\cO \in \phi\times \phi} (c_{\phi\phi\cO})^2 G_{\Delta_\cO}(z)~,
\ee
where we sum over all exchanged operators $\cO$ which appear in the OPE of $\phi\times \phi $.

As   reviewed in the introduction, for any CFT correlators, we can   define the  corresponding Mellin amplitudes \cite{Mack:2009mi}. In the particular case of 1D CFT considered here, we will apply the Mellin techniques following the prescription in \cite{Bianchi:2021piu}. 

For this aim, let us first change variable from $z$ to $t$ via the following simple transformation:  
\be
 z=\frac{t}{1+t} \in[0,1]~, \qquad t=\frac{z}{1-z}\in[0,\infty)~.
\ee

In the new variable,  the four point function   becomes
\be\label{Ghatt}
\widehat G(t) = G\Big(z=\frac{t}{1+t}\Big)~, 
\ee
and the crossing equation \eqref{crossingeq} reduces to
\be\label{Ghatcrossing}
\widehat G(t) = t^{2\Delta_\phi} \;\widehat G\Big(\frac1t\Big) ~.
\ee

The conformal block expansion is now 
\be\label{confdecomp2}
\widehat G(t) =\sum_{\cO \in \phi\times \phi} (c_{\phi\phi\cO})^2\widehat G_{\Delta_\cO}(t)~,
\ee
where
\footnote{Here we used the identity
\be\nonumber
{}_2F_1(a, b; c; z)=(1-z)^{-a} {}_2F_1\Big(a, c-b; c; \frac{z}{z-1}\Big)~.
\ee}
 \be\label{conformalblock2}
\widehat G_\Delta(t)=G_\Delta\Big(z=\frac{t}{1+t}\Big)
=t^\Delta {}_2F_1(\Delta, \Delta; 2\Delta; -t)~.
\ee

We can now define  the celestial Mellin amplitude for the celestial amplitude/correlator \eqref{cGCFT} by performing another Mellin transformation in coordinate   \cite{Bianchi:2021piu}
\be\label{MellinTsf}
\cM(s) =\int_0^\infty dt\, \widehat G(t)\; t^{-1-s} ~.
\ee
This is the main object we would like to study in this paper to understand celestial holography. Note that, the celestial Mellin amplitude is obtained from momentum space amplitudes by performing two times of Mellin transforms \eqref{Gz4} and \eqref{MellinTsf}, one in energy and the other in coordinate. 

To have a meaningful Mellin transform, we need to discuss the convergence of integral in \eqref{MellinTsf}.  We assume that $ \widehat G(t)$ is well behaved for $t\in(0, \infty)$
and the only possible divergence comes from $t\to 0$ and $t\to \infty$. Let us first consider the behavior near $t=0$. As one can see from \eqref{confdecomp2} and \eqref{conformalblock2}, the leading power is $ \widehat G(t)\sim t^{\Delta_\text{lightest}  }$, where $\Delta_\text{lightest}$ is the dimension of the ``lightest'' exchanged operator, namely $\Real \Delta_\text{lightest} =\min_{\cO\in \phi\times \phi} \Real\Delta_\cO$. 
\footnote{Here we consider $\Real\Delta_\cO$ instead of  $\Delta_\cO$ because the   operators in CCFT generally have complex scaling dimensions. }
 Therefore, the requirement of convergent integral near 0 leads to the condition $\Real\Delta_\text{lightest} >\Real s$. Using crossing symmetry \eqref{Ghatcrossing}, we learn that near infinity $t\to\infty$,  $ \widehat G(t)\sim t^{2\Delta_\phi-\Delta_\text{lightest}  }$. In order to have a convergent integral there, we similarly find 
$\Real(2\Delta_\phi-\Delta_\text{lightest}) <\Real s$. Therefore, the Mellin integral  \eqref{MellinTsf} converges if 
\be
\Real(2\Delta_\phi-\Delta_\text{lightest} )<\Real s<\Real\Delta_\text{lightest} ~.
\ee

So to have a well-defined   Mellin transform \eqref{MellinTsf} for some value of $s$, we must require  $\Real\Delta_\text{lightest} >\Real\Delta_\phi$.  Equivalently, this means that all exchanged operators should satisfy: 
\be\label{conditionO}
\forall \cO\in \phi \times \phi~, \qquad \Real\Delta_\cO>\Real \Delta_\phi ~. 
\ee

 Throughout this paper, for simplicity, we will assume the condition  \eqref{conditionO} is always satisfied unless specified explicitly. Physically, we can imagine that $\Delta_\phi$ is a free parameter in the complex plane. If the physical exchanged operators have scaling dimension bounded from below, then we expect there always   exists a range of  $\Delta_\phi$  such  that \eqref{conditionO} holds (at least when the bound is  independent of  $\Delta_\phi$). We can discuss the celestial Mellin amplitude for this range of 
  $\Delta_\phi$, and then analytically continue the final results to the whole complex plane for $\Delta_\phi$. In case the condition \eqref{conditionO} is violated, it is also possible to do Mellin transform by  performing some subtractions and analytic continuations. This has been considered in \cite{Penedones:2019tng,Bianchi:2021piu}.

Given the  celestial Mellin amplitude, we can also compute the celestial amplitude/correlator by performing the inverse Mellin transform 
\be
\widehat G(t) =\int_C\frac{ds}{2\pi i }\; \cM(s) \; t^s~,
\ee
for a suitable choice of contour $C$ from $-i\infty$ to $i\infty$. Since we assume the condition \eqref{conditionO}, any contour is valid as long as $\Real(2\Delta_\phi -\Delta_\text{lightest}  )<\Real s<\Real \Delta_\text{lightest} $.
In case this assumption \eqref{conditionO} does not hold, one needs to deform the counter following the prescription in \cite{Bianchi:2021piu}.

Since we are interested in the celestial correlators in CCFT, we can find a more direct expression of celestial Mellin amplitudes in terms of momentum space amplitudes.  Inserting \eqref{Gz4} into  \eqref{MellinTsf}, we obtain 
 \beqn
 \cM(s) &=&
  16^{-\Delta_\phi}  \int_{-\infty} ^0 d\sfs \int_{-\sfs}^\infty d\sft \; \frac{(-\sfs)^ {2\Delta_\phi-s-\frac32}
 (\sfs+\sft)^ {s-\frac32}}{\sqrt{\sft}} A(\sfs,\sft) 
\\&=&
 16^{-\Delta_\phi} \int_{-\infty} ^0 d\sfs\, d\sfu\; \frac{(-\sfs)^ {2\Delta_\phi-s-\frac32}
 (-\sfu)^ {s-\frac32}}{\sqrt{-\sfs-\sfu}} A(\sfs,\sfu)~,
 \label{MellinSca}
 \eeqn
 where we used $\sfs+\sft+\sfu=0$ and $A(\sfs,\sft)=A(\sfs,\sfu)$ \eqref{Astu}. 
So the celestial Mellin amplitude looks like, but not exactly,  the double Mellin transforms of scattering amplitude in two Mandelstam variables.
 
\subsection{Mellin block expansion}
In this subsection, we would like to discuss the Mellin block expansion of celestial Mellin amplitudes. Some   of the results here have been discussed in  {\cite{Bianchi:2021piu}}, but we will derive  the key and new formula \eqref{OPEcoeff} which relates the OPE coefficients and residues of celestial Mellin amplitudes.

Essentially, we would like to perform the Mellin transform \eqref{MellinTsf} on both sides of  \eqref{confdecomp2} and then obtain  the    block expansion for celestial Mellin amplitude. However, we will do it  in an indirect but illuminating way. 

Using crossing symmetry \eqref{Ghatcrossing}, it is easy to   see that
\be\label{Mscrossing}
\cM(s) =\int_0^\infty dt\, \widehat G(1/t) t^{2\Delta_\phi -1-s}
=\int_0^\infty dr\, \widehat G(r ) r^{s-2\Delta_\phi -1 }
=\cM(2\Delta_\phi-s)~.
\ee
This is the crossing symmetry for celestial Mellin amplitudes. It can also be seen obviously from \eqref{MellinSca} by exchanging  $\sfs\leftrightarrow\sfu$ and using $A(\sfs,\sfu)=A(\sfu,\sfs)$.

This further suggests us to split the integration range in  \eqref{MellinTsf}  into two parts and consider 
\be \label{Mstwo}
\cM(s) =\int_0^1 dt\, \widehat G(t) t^{-1-s}+\int_1^\infty dt\, \widehat G(t) t^{-1-s}
=\int_0^1 dt\, \widehat G(t) t^{-1-s}+\int_0^1 dr\, \widehat G(r ) r^{s-2\Delta_\phi -1 }~.
\ee
Therefore, we can rewrite the celestial Mellin amplitude as
\be\label{Mhalf}
\cM(s) =\cM_{\frac12}(s)+\cM_{\frac12}(2\Delta_\phi-s)~,
\ee
where the half Mellin amplitude is
\be\label{halfMs}
\cM_{\frac12}(s) =\int_0^1 dt\, \widehat G(t) t^{-1-s}~.
\ee
The representation in \eqref{Mhalf} has the big advantage that it makes the crossing symmetry \eqref{Mscrossing} manifest. 

We can now plug  in the conformal block expansion \eqref{confdecomp2} into the \eqref{halfMs}. 
In particular,   the conformal block itself \eqref{conformalblock2} gives rise to the half Mellin block
\be\label{Fdeltas2}
B_\Delta(s)=\int_0^1 dt\, \widehat G_\Delta(t)\; t^{-1-s}
=\frac{ {}_3F_2 (\Delta, \Delta, \Delta-s; 2\Delta, \Delta-s+1; -1) }{\Delta-s}~,
\ee
where the integral is performed  on   condition that $\Real s<\Real \Delta$. 
 
 As a result, we arrive at the block expansion of celestial Mellin amplitude
 \be\label{Msexpansion}
\cM(s) =\sum_\Delta C_\Delta    M_\Delta(s)  
=\sum_\Delta C_\Delta  
  \frac{ {}_3F_2 (\Delta, \Delta, \Delta-s; 2\Delta, \Delta-s+1; -1) }{\Delta-s}
  +\Big\{s\to 2\Delta_\phi-s \Big\}~,
\ee
where where we sum over all exchanged operators $\cO_\Delta\in \phi\times \phi$ and $C_\Delta\equiv (c_{\phi\phi\cO_\Delta})^2$.  Here the Mellin block  is given by
\be\label{MdeltaF}
   M_\Delta(s)= B_\Delta(s) +B_\Delta(2\Delta_\phi-s) ~.
\ee
The nice property of Mellin blocks is that by construction they are manifestly crossing symmetric 
\be
   M_\Delta(s)=   M_{ \Delta}(2\Delta_\phi-s)~.
 \ee
 
From \eqref{Msexpansion}\eqref{Fdeltas2}, it is obvious to see that  $s=\Delta$ is the pole of celestial Mellin amplitude.  This is   an interesting feature. However, a crucial point is that there are also other poles in \eqref{Msexpansion}. To illustrate this, we first consider    the case $\Delta=1$. Then \eqref{Fdeltas2} takes the following explicit form
\be
B_1(s)=
\frac{\psi ^{(0)}\left(1-\frac{s}{2}\right)}{2 s}+\frac{-\psi ^{(0)}\left(\frac{1}{2}-\frac{s}{2}\right)+\gamma_E +\psi ^{(0)}\left(\frac{1}{2}\right)}{2 s}~,
\ee
where $\psi^{(0)}(z) =\Gamma'(z)/\Gamma(z)$ is the polygamma function, and $\gamma_E=0.577215\cdots $ is the Euler-Gamma constant. It is easy to  show that this function $B_1(s)$ has poles at all positive integers: 
\be\label{F1s}
\lim_{s\to n} B_1(s)\sim \frac{(-1)^n/n}{s-n}, \qquad n =1,2,3, \cdots~.
\ee

Actually, we can obtain similar explicit expressions for arbitrary $B_\Delta$  just by employing the     defining series of generalized hypergeometric functions: 
\be\label{hyperGfcn}
{}_p F_q(a_1, \cdots, a_p; b_1, \cdots, b_q; x)=\sum_{n=0}^\infty \frac{(a_1)_n \cdots (a_p)_n}{(b_1)_n \cdots (b_q)_n}\frac{x^n}{n!}~, \quad \quad  (a)_n=a(a+1) \cdots(a+n-1)=\frac{\Gamma(a+n)}{\Gamma(a)}~.
\ee
Note that the infinite series would terminate and reduce to a polynomial if $a_i\in \mathbb Z_{\le 0}$ is a non-positive integer.
 
 Using \eqref{hyperGfcn}, we then find  the following expansion for  the half Mellin block   \eqref{Fdeltas2}   
\be\label{Fdeltas}
B_\Delta(s)=\sum_{n=0}^\infty (-1)^n \frac{(\Delta)_n^2}{n! (2\Delta)_n} \frac{1}{\Delta+n-s}~,
\ee
and it has simple poles at $s=\Delta+k$ with residues 
\be
\Res_{s=\Delta+k}B_\Delta(s)=(-1)^{k+1} \frac{(\Delta)_k^2}{k! (2\Delta)_k}~, \qquad k=0,1, 2\cdots~.
\ee
Setting $\Delta=1$ leads to the equation \eqref{F1s}, as expected. 

Collecting all   results together, we obtain the following equation for celestial Mellin amplitude
\be\label{MsExp}
\cM(s) =\sum_\Delta C_\Delta    M_\Delta(s) 
=\sum_\Delta  \sum_{n=0}^\infty   \frac{C_\Delta}{\Delta+n-s}   \frac{(-1)^n(\Delta)_n^2}{n! (2\Delta)_n} 
+\Big\{  s\to 2\Delta_\phi-s\Big\}~.
\ee
So celestial Mellin amplitudes   have  only simple poles, whose positions and residues encode the information of OPE data.

Let us consider the residue at the  pole $s=s_*$.    Note that due to crossing symmetry  \eqref{Mscrossing},  a pole at $s_*$ implies a mirror pole at $2\Delta_\phi-s_*$ with opposite residue.  Therefore, without loss of generality, we can  assume that $\Real s_*>\Real \Delta_\phi$.  The  residue at the pole $s_*$ is then given by
\beqn\label{2ndtermcousin}
\Res_{s=s_*} \cM(s) &=& \sum_\Delta  \sum_{n=0}^\infty   C_\Delta \frac{(-1)^{n+1}(s_*-n)_n^2}{n! (2s_*-2n)_n} \Theta\Big(\Delta =s_*-n \Big) 
-\Big\{ s_*\to 2\Delta_\phi-s_*\Big\}
\\&=&
 \sum_{n=0}^\infty    \frac{(-1)^{n+1}(s_*-n)_n^2}{n! (2s_*-2n)_n} 
\sum_{\Delta_\cO=s_*-n}(c_{\phi\phi \cO})^2~,
\eeqn
where the step function $\Theta=1$ if $\Delta =s_*-n $ is satisfied and vanishes otherwise.  Note that in the second equality, we dropped the mirror contribution from the second term in \eqref{2ndtermcousin}  because there the operators have dimension $\Real \Delta =\Real(2\Delta_\phi-s_*-n)\le\Real( 2\Delta_\phi-s_*) <\Real\Delta_\phi$, thus violating our assumption \eqref{conditionO}.

Obviously, for the consideration of pole at $s_*$, the residue only gets contribution from  operators whose scaling dimensions differ from $s_*$ by an integer. 
Together with our assumption \eqref{conditionO}, this implies that  all the operators $\cO_l$ contributing to the residue   have scaling dimension of the form $\Delta_l\equiv\Delta_0+l$ with $l \in \mathbb N$.
Obviously, $s_*$ can only differ from $\Delta_0$ by an integer, so we can take $s_*=\Delta_0+k$ with $k \in \mathbb N$. These considerations then lead  to
\beqn
\Res_{s=\Delta_0+k} \cM(s)&=& \sum_{n=0}^\infty    \frac{(-1)^{n+1}(\Delta_0+k-n)_n^2}{n! (2\Delta_0+2k-2n)_n} 
\sum_{\Delta_l=\Delta_0+k-n}(c_{\phi\phi \cO_l})^2
\\&=&
 \sum_{n=0}^\infty    \frac{(-1)^{n+1}(\Delta_0+k-n)_n^2}{n! (2\Delta_0+2k-2n)_n} 
 (c_{\phi\phi \cO_{k-n}})^2
 \\&=&
 \sum_{l=0}^k    \frac{(-1)^{k-l +1}(\Delta_0+l)_{k-l}^2}{(k-l)! (2\Delta_0+2k)_{k-l }} 
 (c_{\phi\phi \cO_{l}})^2~.
 \label{MellinResidueBlock}
\eeqn
This equation takes the form  of $F_k=\sum_{l=0}^k  M_{kl}C_l$, where  \be\label{Mmatrix}
F_k=\Res_{s=\Delta_0+k} \cM(s)~, \qquad
C_l=(c_{\phi\phi \cO_{l}})^2~,\qquad
M_{kl}=
\frac{(-1)^{k-l +1}(\Delta_0+l)_{k-l}^2}{(k-l)! (2\Delta_0+2k)_{k-l }}~ .  
\ee 
They can be further regarded as the entries of infinite dimensional vectors  $\bm F, \bm C$ and matrix $\bm M$, respectively. Then \eqref{MellinResidueBlock} can be compactly written as $\bm F=\bm M\bm C$. Once we know the celestial Mellin amplitude, we can compute the residues at all the poles, namely the vector  $\bm F$. Then the OPE coefficients are encoded in the vector   $\bm C=\bm M^{-1}\bm F=\bm N \bm F$, where $\bm N$ is the inverse matrix of $\bm M$. We can find the matrix $\bm M$ and inverse explicitly  at the first few orders: 
\footnote{Note that  although $\bm M$ is an infinite dimensional matrix, its  inverse is   actually  easy to compute because $\bm M$ is lower triangular. We can write $\bm M =-(\bm I+\bm T)$ where $\bm I$ is the identity matrix, then $\bm M^{-1}=- \sum_{\ell=0}^\infty (-1)^\ell \bm T^\ell$. More explicitly $(\bm M^{-1})_{kl}= -\delta_{kl}+ \sum_{ k>j_1>j_2\cdots >j_f>l}(-1)^f    T_{k j_1}
   T_{j_1 j_2}   T_{j_2j_3}\cdots T_{j_f l}.
$ }
\be
\bm M=\left(
\begin{array}{cccc}
 -1 & 0 & 0&\cdots \\
 \frac{\Delta_0}{2} & -1 & 0 &\cdots \\
 -\frac{\Delta_0 (\Delta_0+1)^2}{8 \Delta_0+4} & \frac{\Delta_0+1}{2} & -1&\cdots \\
\vdots &\vdots &\vdots &\ddots\\
\end{array}
\right), \qquad
\bm N=\bm M^{-1}=\left(
\begin{array}{cccc}
 -1 & 0 & 0&\cdots \\
 -\frac{\Delta_0}{2} & -1 & 0 &\cdots\\
- \frac{\Delta_0^2 (\Delta_0+1) }{8 \Delta_0+4}     & -\frac{\Delta_0+1}{2} & -1&\cdots \\
\vdots &\vdots &\vdots &\ddots\\
\end{array}
\right)~.
\ee
Note that both $\bm M$ and its inverse $\bm N $ are lower triangular matrices. It is natural to ask whether one can find the analytic closed expression for $\bm N$  explicitly.  
Indeed, after some trivial and error, we find the explicit expression for the inverse:  
\be\label{Minv}
N_{kl}= \frac{(-1)^{-\left\lfloor \frac{k-l}{2}\right\rfloor } 2^{-\left\lfloor \frac{ (k-l-2)}{2}\right\rfloor-\left\lfloor \frac{1}{2} (k-l+1)\right\rfloor } }{(2 \Delta_0+2 k-3) \; (k-l)!}
 \frac{\Gamma (\Delta_0+k) \;  \Gamma \left(\Delta_0+l+\left\lfloor \frac{k-l}{2}\right\rfloor \right)}{  \Gamma (\Delta_0+l)^2 \; \left(-\Delta_0-k+\frac{5}{2}\right)_{\left\lfloor \frac{ (k-l-2)}{2}\right\rfloor}}~, 
 \ee
 where the floor function $\left\lfloor x \right\rfloor$ gives the  greatest integer less than or equal to   $x$.   Surprisingly, we find that the expression can be simplified dramatically     
 \be\label{Nmatrix}
 N_{kl} 
=  -\frac{\Gamma ( \Delta_0+k )^2 \Gamma ( 2 \Delta_0+k+l -1)}{\Gamma ( 2 \Delta_0 +2 k-1) \Gamma (k-l+1) \Gamma ( \Delta_0+l )^2}~.
 \ee
Furthermore, one can   show that this matrix $\bm N$ is indeed the inverse of $\bm M$ in \eqref{Mmatrix}. We present the proof in Appendix~\ref{MNinverse}.
 
 As a result, we find that the OPE coefficients are given by 
\beqn 
 (c_{\phi\phi \cO_{k }})^2&=&\sum_{l=0}^k N_{kl}\Res_{s=\Delta_0+l} \cM(s) 
\nonumber\\&=&
-\frac{\Gamma ( \Delta_0+k )^2  }{\Gamma ( 2 \Delta_0 +2 k-1)  }
\sum_{l=0}^k  \frac{ \Gamma ( 2 \Delta_0+k+l -1)}{ \Gamma (k-l+1) \Gamma ( \Delta_0+l )^2}
 \Res_{s=\Delta_0+l} \cM(s) ~.
 \label{OPEcoeff}
\eeqn
This gives the explicit formula to compute the OPE coefficients, once we know the residues of celestial Mellin amplitudes.


 \subsection{Relation to coefficient functions from inversion formulae} \label{inversionForm}

In the previous subsection, we have shown that   the poles and residues of celestial Mellin amplitudes just encode the information of OPE data in CCFT. This is reminiscent of the coefficient  functions in CFT which can be calculated using inversion formulae. In this subsection, we would like to relate the celestial Mellin amplitudes  with coefficient  functions.
 
  For 1D conformal group $SL(2, \mathbb R)$, the complete set of wave function  includes both the principal  continuous series $\Delta\in\frac12 +i \mathbb R_+$ and discrete series $\Delta\in 2 \mathbb Z_{>0}$; the latter is absent in  higher dimensions. 
Given a four-point  correlator $G(z)$ in 1D CFT, one can define the following coefficient functions  for  the principal and discrete series conformal partial waves
\beqn\label{eq:euclInv0}
I_\Delta &=&  \int_{-\infty}^{\infty} dz \;z^{-2}\,\Psi_{\Delta}(z) \, {G}(z)~,
\quad\quad\quad\textrm{for }\Delta=\frac{1}{2}+i r~,\quad r\in\mathbb{R}_+~,
\\
\widetilde{I}_m &=&  \int_{-\infty}^{\infty} dz \;z^{-2}\,\Psi_{m}(z)\, {G}(z)~,
\quad\quad\quad\textrm{for }m\in2\mathbb{Z}_{>0}\,.
\label{eq:euclInv}
\eeqn 
These are the Euclidean inversion formulae.  Here $\Psi_\Delta $ is the conformal partial wave  which we  review  in Appendix~\ref{AppInversion}, see \eqref{Psi0pm}. The set of $\Psi_\Delta$ including both principal and discrete series  forms a complete and orthogonal basis of wave functions for the conformal group $SL(2, \mathbb R)$. 

In terms of the coefficient functions $I_\Delta , \widetilde{I}_m$, the   four-point function can   be expanded in the complete set and then rewritten as   follows  \cite{Mazac:2018qmi}
\be 
 {G} (z) = \int_ {\frac{1}{2} - i\infty}^{\frac{1}{2} + i\infty} \frac{d\Delta}{2\pi i}\frac{I_{\Delta}}{2K_{\Delta}}G_{\Delta}(z) + \sum\limits_{m\in2\mathbb{Z}_{>0}}\frac{\Gamma (m)^2}{2 \pi ^2 \Gamma (2 m-1)}\widetilde{I}_m G_{m}(z)~, \qquad 0<z<1~.
\label{eq:opeFromIs}
\ee
 We can deform the contour toward the right half plane and pick  up various poles.   Compared to \eqref{confdecomp}, we have
\be\label{OPEcoef}
(c_{\phi\phi \cO})^2=
-\Res_{\Delta=\Delta_\cO}\frac{I_{\Delta}}{2K_{\Delta}}
+ \sum\limits_{m\in2\mathbb{Z}_{>0}}
\frac{\Gamma (m)^2}{2 \pi ^2 \Gamma (2 m-1)}\widetilde{I}_m  \;\delta_{m , \Delta_\cO}~.
\ee
Therefore, up to the contribution from discrete series,  the position of simple poles  give  precisely the   scaling dimensions of exchanged operators,  and the residues  there are just the   OPE coefficient squares.

In addition to the Euclidean   inversion formulae, one can also use Lorentzian inversion formulae  to compute   $I_\Delta , \widetilde{I}_m$.  
The Lorentzian inversion formula for the discrete series coefficient function is \cite{Simmons-Duffin:2017nub}
\be\label{LorentzInv}
\widetilde{I}_m = \frac{4\Gamma(m)^2}{\Gamma(2m)}\int _{0}^{1} dz \; z^{-2}G_{m}(z)\dDisc\!\left[ {G}(z)\right]~,\quad\qquad m \in 2 \mathbb Z_{>0}~,
\ee
where $\dDisc[G(z)]$ is the double discontinuity which we review in \eqref{doubleDiscon}. 
The above formula is valid for any physical four-point function satisfying $ {G}(z) =  {G}\!\left(\mbox{$\frac{z}{z-1}$}\right)$.  This formula  provides a particular analytic continuation of $\widetilde{I}_\Delta$ beyond the discrete series $\Delta\in 2\mathbb{Z}_{>0}$. However, the resulting analytically continued $\widetilde{I}_\Delta$  in general does not necessarily agree with the principal series coefficient function $I_{\Delta}$.

 There is also  a  Lorentzian inversion formula for the principal series coefficient function    \cite{Mazac:2018qmi}
 \be
I_\Delta=2\int_0^1 dz\; z^{-2} H_\Delta(z) \dDisc [G(z)]~.
\ee 
Here the inversion kernel $H_\Delta(z) $ needs to satisfy various properties. The  closed explicit expression of $H_\Delta(z) $ is in general not known, except for some special cases. 

Since both coefficient functions $I_\Delta, \widetilde I_m$ and   celestial Mellin amplitudes encode the OPE data in terms of their simple poles and residues, we expect that they should be related to each other. Very straightforwardly, we can apply Mellin transform \eqref{MellinTsf} on both sides of  \eqref{eq:opeFromIs}, and find 
\be
\cM(s) = \int _{\frac{1}{2} - i\infty}^{\frac{1}{2} + i\infty}\!\frac{d\Delta}{2\pi i}\frac{I_{\Delta}}{2K_{\Delta}}  \frac{\Gamma(s)^2 \Gamma(2\Delta) \Gamma(\Delta-s)}{\Gamma(\Delta)^2 \Gamma(\Delta+s)}
+ \sum\limits_{n=1}^\infty   \frac{\Gamma(s)^2 \Gamma(4n) \Gamma(2n-s)}{2 \pi ^2 \Gamma (4 n-1) \Gamma(2n+s)}\widetilde{I}_{2n} ~,
\label{eq:opeFromIs22}
\ee
where we used the following Mellin transform of conformal block 
   \be \label{MellinTsfOfConfBlock}
   \widehat G_\Delta(t) \quad  \xrightarrow{\eqref{MellinTsf}} \quad
  \frac{\Gamma(s)^2 \Gamma(2\Delta) \Gamma(\Delta-s)}{\Gamma(\Delta)^2 \Gamma(\Delta+s)}~, \qquad
  \Real \Delta >\Real s>0~.
  \ee

Alternatively,  we can split the Mellin integral \eqref{MellinTsf} into two regions following  \eqref{Mstwo}, and derive the following  illuminating   and manifestly crossing-symmetric equation
\be
\cM(s) =\int_{\frac{1}{2} - i\infty}^{\frac{1}{2} + i\infty}\!\frac{d\Delta}{2\pi i}\frac{I_{\Delta}}{2K_{\Delta}}  M_\Delta(s)
+ \sum\limits_{m\in2\mathbb{Z}_{>0}}\frac{\Gamma (m)^2}{2 \pi ^2 \Gamma (2 m-1)}\widetilde{I}_m M_{m}(s)~,
\label{CMAvsOPEcoff}
\ee
where $M_\Delta$ is the Mellin block \eqref{MdeltaF}\eqref{Fdeltas2}.
From \eqref{Fdeltas} and \eqref{MdeltaF}, it is easy to see that a pole $\Delta_*$ in $I_\Delta/K_\Delta$ induces an infinite tower of poles  $\Delta_*+k$ together with their mirrors $2\Delta_\phi-\Delta_*-k$  in $\cM(s)$, with $k\in \mathbb N$. There are also additional poles at $m+k$ and $2\Delta_\phi-m-k$ coming from discrete series, with  $k\in \mathbb N, m\in2\mathbb{Z}_{>0}$. Note that due to possible cancellation between principal and discrete series, some of  the poles at positive even integers may be absent.

\subsection{Regge behavior} \label{ReggeBH}
 
  In this subsection, we discuss the      behavior of celestial Mellin amplitudes in the Regge limit, which  is a very interesting limit in standard unitary CFT and  has a close connection with chaos.
  
 In 1D CFT, we have the $s$-channel limit, $x_1\to x_2$, and the $t$-channel limit $x_2\to x_3$. But strictly speaking, there is no notion of $u$-channel limit   as it is impossible to bring $x_1$ close to $x_3$ without  crossing $x_2$. However, from higher dimensions, one can infer that    the $u$-channel Regge limit is $z\to i \infty$, more precisely arbitrary points at infinity on the upper or lower half complex plane excluding the real axis.
For concreteness, we consider $z=\frac12 + i \xi$ with $\xi \to \infty $.
Then one can show that in standard  unitary 1D CFTs, the  four-point functions      are bounded in the Regge limit   \cite{Mazac:2018qmi}
\be\label{Reggebounded}
\Big|\widetilde G\Big(\frac12 + i \xi\Big)\Big |  <\infty~   \qquad \text{as} \qquad \xi \to \infty ~,
\ee
where $ \widetilde G(z)=z^{-2\Delta_\phi} G(z)
 =\widetilde G(1-z)$.

We would like to translate \eqref{Reggebounded} into Mellin amplitude. For  convenience, we    normalize  the Mellin amplitude  following \cite{Bianchi:2021piu}
\be
{\bm \cM}(s) =\frac{\cM(s)}{\Gamma(s) \Gamma(2\Delta_\phi-s)}~.
\ee
If the Regge limit boundedness \eqref{Reggebounded} holds, then the normalized Mellin amplitude  satisfies  \cite{Bianchi:2021piu}
 \footnote{ Here one needs to assume the absence of   Stokes phenomena, otherwise the bound is ${\bm \cM}(c+i\eta) =\sO(|\eta|^0)$, as $ \eta\to\infty$ for proper $c$ to ensure convergence. 
}
\be\label{Reggebounded2}
{\bm \cM}(s)=\sO(|s|^0)~, \qquad  | s| \to\infty~.
\ee

Since we are interested in 1D CCFTs which are not   standard unitary CFTs, there is no guarantee for the Regge boundedness behavior.  Actually, the meaning of Regge limit is even not clear at all, and it would be very interesting to understand further. 
Nevertheless, one can show that   if there exits $n$  such that
\be\label{regg1}
 G(z) =\sO(z^{2\Delta_\phi+n})~ , \qquad z \to \frac12 +i\infty~,
\ee
then the normalized Mellin amplitude behaves as  
\be\label{regg2}
{\bm \cM}(s) =\sO(|s|^n)~, \qquad | s| \to\infty~.
\ee

\section{An example} \label{CMAexample}

In this section, we   illustrate many features   of celestial Mellin amplitude in a concrete scalar model in 3D. We will first present the model, compute the celestial amplitude and its conformal block expansion. Then we  will calculate the celestial Mellin amplitude and its Mellin block expansion, which is shown to agree  with the conformal block expansion. Finally, we will use inversion formulae to compute the coefficient functions in order to compare it with celestial Mellin amplitude. 

\subsection{A simple model in 3D}
Let us consider a simple model in (2+1)D consisting of  one real massless field $\phi$ and one real massive scalar field $\sigma$ with mass $m$. They interact  via the cubic vertex $g \phi^2 \sigma$. This model was studied in \cite{Lam:2017ofc}. It is easy to compute the tree-level four massless particle scattering amplitude  due to the massive particle exchange: 
\be\label{Ast}
A(\sfs,\sft) =-\frac{g^2}{\sfs-m^2}-\frac{g^2}{\sft-m^2}-\frac{g^2}{\sfu-m^2}~,
\ee 
where we have ignored the $i\epsilon $ factor.  Plugging into \eqref{Gz4},  one can then obtain  the corresponding celestial amplitude \cite{Lam:2017ofc}
\be\label{cAmpeg}
G (z) =G^{(0)}(z) =\mathcal N \frac{z}{\sqrt{1-z}} \Bigg[1+e^{\pi i \alpha}z^\alpha+\Big(\frac{z}{1-z} \Big)^\alpha \Bigg]~, \qquad 0<z<1~,
\ee
with 
\be\label{alphaN}
\alpha \equiv 2\Delta_\phi -\frac32~, \qquad \mathcal N =\frac{g^2 \pi m^{4\Delta_\phi-5}}{2^{4\Delta_ \phi}\cos (2\pi \Delta_\phi)}~.
\ee
Here we   have taken 1,3 incoming, 2,4 outgoing.    Note that there is no delta function because we are considering scattering in three dimensional spacetime. 
 For simplicity, we will set $\cN=1$.

\subsection{Conformal block expansion}

We would like to perform the conformal block expansion for the celestial correlator \eqref{cAmpeg}.  Such a decomposition has been studied for a similar model in
\cite{Garcia-Sepulveda:2022lga}. 

The conformal block expansion can be done with the help of the following identity \cite{Hogervorst:2017sfd}: 
 \be\label{usefulFormula}
 \frac{z^p}{(1-z)^q}=\sum_{n=0}^\infty  \sfC_{p,q}(n) G_{p+n} (z)~, \qquad 0<z<1~,
 \ee
 where $G_\Delta$ is the conformal block in \eqref{confblocks} and 
 \be\label{cpqn}
 \sfC_{p,q}(n)=\frac{(p)_n^2}{n! (2p+n-1)_n} {}_3F_2(p-q, 2p+n-1, -n; p,p; 1)~, \qquad
 (q)_n=\frac{\Gamma(q+n)}{\Gamma(q)}~.
 \ee

 Using this identity \eqref{usefulFormula}, it is easy to   find the conformal block expansion  of celestial amplitude \eqref{cAmpeg}
 \be\label{cfblockexp}
 G(z) =\sum_{k=0}^\infty\frac{16^k \Gamma \left(k+\frac{1}{2}\right)^2}{\Gamma \left(\frac{1}{2}-k\right)^2 \Gamma (4 k+1)}
G_{2k+1}(z) 
 +\sum_{n=0}^\infty  \Big(1+(-1)^n e^{\pi i \alpha} \Big)\sfC_{1+\alpha, \frac12+\alpha} (n)\, G_{n+1+\alpha}(z) ~,
 \ee
 where we used the property   
 \be\label{cvanish}
 \sfC_{1,\frac12}(2k+1) =0~, \qquad \sfC_{1+\alpha, \frac12}(n) =(-1)^n \sfC_{1+\alpha\,, \frac12+\alpha}~.
 \ee
 
 Then it is easy to read off the scaling dimension of exchanged operators 
 \be\label{cOexchange}
 \Delta_\cO=2n+1~, \qquad
 \Delta _\cO =n+ 2\Delta_\phi-\frac12~, \qquad n \in \mathbb N~.
 \ee
 
 Note that operators with positive even integer scaling dimensions are absent in \eqref{cOexchange}. Furthermore, in  the conformal block expansion \eqref{cfblockexp}, the squared OPE coefficients   are complex (unless $\Real\alpha \in \mathbb Z$), which is thus a signature of non-unitarity.

\subsection{Celestial Mellin amplitude  } 
 
We can calculate the Mellin transform \eqref{MellinTsf}  of celestial correlator \eqref{cAmpeg} and obtain the following celestial Mellin amplitude
\be\label{CAMeg}
\cM(s) =\frac{\Gamma(1-s) \Gamma(s-\frac12)}{\sqrt \pi}
+\frac{\Gamma(s+1-2\Delta_\phi) \Gamma( 2\Delta_\phi-s-\frac12)}{\sqrt \pi}
+ {i e^{2\pi i \Delta_\phi} }  \frac{\Gamma(s-\frac12) \Gamma( 2\Delta_\phi-s-\frac12)}{\Gamma(2\Delta_\phi-1)}~, \qquad
\ee
provided that  $\frac12 <\Real s \,,\; \Real( 2\Delta_\phi-s)<1$, and hence $\frac12<\Real \Delta_\phi<1$.   Without further justification, we will analytic continue the results on the complex plane, so \eqref{CAMeg} will be assumed to be  valid for all $s\in \mathbb C$.  In the above evaluation, we have used the following equation  for Mellin transform 
 \beqn\label{zpqTsf}
  \frac{z^p}{(1-z)^q}=t^p (1+t)^{q-p}
 \quad& \xrightarrow{\eqref{MellinTsf}}&\quad
  \frac{\Gamma(p-s)\Gamma(s-q)}{\Gamma(p-q)}, \qquad
  \Real p >\Real s>\Real q~.
 \eeqn

 Alternatively, the celestial Mellin amplitude can also be obtained directly from momentum space amplitude. Inserting \eqref{Ast} into \eqref{MellinSca}, we get exactly   \eqref{CAMeg}, including the factor $\mathcal N$ in \eqref{alphaN}. 
\footnote{When performing the integrals, the following two identities turn  out to be useful
\be
 \int_0^ \infty dS dU\; \frac{S^a U^b}{\sqrt{S+U}}\frac{1}{S+M}
 =- \frac{M^{\frac12+a+b}\sqrt{\pi}}{\cos(\pi(a+b))} \Gamma(-\frac12-b) \Gamma(1+b)~, 
 \nonumber
 \ee
 
 \beqn
 \int_0^ \infty dS dU\; \frac{S^a U^b}{\sqrt{S+U}}\frac{1}{S+U+M}
 &=&
  \int_0^ \infty dX \int_0^1  dx\; \frac{X^{a+b+\frac12} x^a (1-x) ^b}{ X+M}
\nonumber \\&=&
 - \frac{M^{\frac12+a+b} {\pi}}{\cos(\pi(a+b))} \frac{ \Gamma(1+a ) \Gamma(1+b)}{\Gamma(2+a+b)}~.
 \nonumber
  \eeqn
 }
  
 Now we would like to see the behaviors of celestial amplitude  \eqref{cAmpeg} and celestial Mellin amplitude    \eqref{CAMeg}    in the Regge limit discussed in   section~\ref{ReggeBH}.
For the  celestial amplitude  \eqref{cAmpeg}, the Regge behavior \eqref{regg1} is easy to see 
\be
G (z)\sim z^{\frac12}~, \quad 
z^{2\Delta_\phi-1 }~, \qquad \text{ as} \quad z\to i\infty~.
\ee
For the  (normalized)  celestial Mellin amplitude  \eqref{CAMeg}, the Regge behavior \eqref{regg2} is also easy to find~\footnote{Here we use the Stirling's approximation for Gamma function,   
$
\Gamma(x) \sim \sqrt{2\pi} x ^{x-\frac12} e^{-x}   \Big( 1+\sO( 1/ x)\Big),  
$
as $|x|\to \infty $ at fixed $|\arg(x)|<\pi$. Thus up to an order one factor,  we have
$
\Gamma(a+x) \Gamma(b-x)\sim 
  x^{a+b-1}  
$.}
\be
{\bm \cM}(s) \equiv\frac{\cM(s)}{\Gamma(s) \Gamma(2\Delta_\phi-s)}\sim s^{\frac12-2\Delta_\phi}~, \quad 
s^{  -1 } ~, \qquad \text{ as} \quad |s|\to  \infty~.
\ee

Therefore, we see that the    (normalized)   celestial Mellin amplitude does not satisfy the Regge boundedness condition \eqref{Reggebounded2}, unless $\Real \Delta_\phi>\frac14$.
 
\subsection{Mellin block expansion  }
 We can easily check   the crossing invariance $\cM(s) =\cM(2\Delta_\phi-s)$ of celestial Mellin amplitude \eqref{CAMeg}. 
Consequently, the  pole   $s_*$ and its mirror    $2\Delta_\phi-s_*$ alway appear together in pairs. 

From \eqref{CAMeg}, it is   easy to read  off all the simple poles 
\be\label{spole1}
s=k+1~, \; -k+\frac12~, \; 2\Delta_\phi-(k+1)~, \; 2\Delta_\phi- (-k+\frac12)~, \qquad k=0,1,2,\cdots~.
\ee
For simplicity, we assume that $\Delta_\phi$ is generic, so all these poles do not coincide with each other. 

Here we find the poles of celestial Mellin amplitude, but our actual interest is the operator content in CCFT. Note that not all the poles of celestial Mellin amplitude correspond to the scaling dimension of physical exchange operators. First of all, due to crossing symmetry, the poles at $s=\Delta $ and $s=2\Delta_\phi-\Delta$ actually come from the same physical operators \eqref{MsExp}. Secondly, it is possible that the physical exchange operator are actually not present at the pole of celestial Mellin amplitude due to the nontrivial relation \eqref{OPEcoeff} between OPE coefficients and the residues of the celestial Mellin amplitude. In that case, the residues of the celestial Mellin amplitude contrive in a tricky way, leading   to   vanishing OPE coefficients in  \eqref{OPEcoeff}.  Nevertheless,  the poles of the celestial Mellin amplitude give  the maximal possible set of scaling dimension of operators.  If we further assume \eqref{conditionO},  we should then only keep those poles with $\Real s>\Real\Delta_\phi$, hence resolving the first ambiguity.  In addition, the formula \eqref{OPEcoeff} allows us to compute the OPE coefficients explicitly from residues of the celestial Mellin amplitude.  All operators with non-vanishing OPE coefficients then correspond to the physical exchange operators. 

Focusing the problem at hand \eqref{spole1} and requiring  $\Real s>\Real\Delta_\phi$, we naturally find the subset 
\be\label{spole}
s=k+1~, \; \qquad 2\Delta_\phi+k-\frac12~ , \qquad k=0,1,2,\cdots~.
\ee
Compared to \eqref{cOexchange}, we see that operators with even scaling dimension $s\in 2 \mathbb Z_{>0}$ are actually absent. This is due to the vanishing OPE coefficient \eqref{cvanish}.  Below, we will show that we can compute explicitly all the OPE coefficients from the celestial Mellin amplitude using the formula \eqref{OPEcoeff}. The result agrees with  \eqref{cfblockexp}\eqref{cvanish} obtained from   conformal block expansion. This then automatically implies that those operators with $\Delta\in 2   \mathbb Z_{>0}$ are actually absent.

Let us now compute the OPE coefficients from celestial Mellin amplitude \eqref{CAMeg}.
Using the following formula 
\be
\lim_{x\to k} \Gamma(x+m) =\frac{1}{x-k} \frac{(-1)^{-m-k}}{(-m-k)!}~, \qquad
k+m=0, -1, -2, \cdots~,
\ee
we find the residues of  celestial Mellin amplitude   \eqref{CAMeg}  
\beqn
\Res_{s=k+1}\cM(s) &=&-\Res_{s=2\Delta_\phi-(k+1)}\cM(s) =\frac{(-1)^{k+1}}{k!} \frac{\Gamma(k+\frac12)}{\sqrt \pi}~,
\\
\Res_{s=-k+\frac12}\cM(s) &=&-\Res_{s=2\Delta_\phi-(-k+\frac12)}\cM(s) 
\\&=&
\frac{(-1)^k \Gamma \left(k+\frac{1}{2}\right)}{\sqrt{\pi } k!}+\frac{i e^{2 i \pi  \Delta_ \phi } (-1)^k\; \Gamma (k+2 \Delta_ \phi -1)}{k! \;\Gamma (2 \Delta_ \phi -1)}~.
\eeqn

With these results, we can compute the OPE coefficients using the formula \eqref{OPEcoeff}. 
As one can see from \eqref{spole}, these two infinite towers of poles  take the form $\Delta_0+k$, with $\Delta_0=1$ and $\Delta_0=2\Delta_\phi-\frac12$ respectively.  At very low orders, one can check explicitly that the resulting OPE coefficients computed by {\eqref{OPEcoeff}}  from celestial Mellin amplitude      are consistent with that obtained from conformal block expansion in \eqref{cfblockexp}. 
Actually, we can prove the full equivalence analytically.   Since the conformal block expansion \eqref{cfblockexp} just follows from the identity \eqref{usefulFormula} and \eqref{cpqn}, we only need to show the general case in  {\eqref{usefulFormula}}, whose corresponding ``celestial Mellin amplitude'' is given in \eqref{zpqTsf}.  
The   OPE coefficients can be then computed using \eqref{OPEcoeff}: 
\footnote{ Alternatively, from  \eqref{MellinResidueBlock}   we just need to show
\beqn\label{cMG}\nonumber
 \sum_{l=0}^k    \frac{(-1)^{k-l +1}(p+l)_{k-l}^2}{(k-l)! (2p+2k)_{k-l }} 
\sfC_{p,q}(l)
=\Res_{s=p+k}   \frac{\Gamma(p-s)\Gamma(s-q)}{\Gamma(p-q)}
=\frac{(-1)^{k+1}\Gamma(p-q+k)}{k!\Gamma(p-q)}~.
\eeqn
This can  be proved by performing the Mellin transform on both sides of \eqref{usefulFormula} 
 \be
  \frac{\Gamma(p-s)\Gamma(s-q)}{\Gamma(p-q)}
  =\sum_{n=0}^\infty \sfC_{p,q}(n) 
  \frac{\Gamma(s)^2 \Gamma(2(p+n)) \Gamma(p+n-s)}{\Gamma(p+n)^2 \Gamma(p+n+s)} ~,
   \ee
and then taking residue at pole $s=p+k$. 
 }
\beqn
 (c_{\phi\phi \cO_{k }})^2&=&\sum_{l=0}^k N_{kl}\Res_{s=p+l}   \frac{\Gamma(p-s)\Gamma(s-q)}{\Gamma(p-q)}
\\&=&
-\frac{\Gamma ( p+k )^2  }{\Gamma ( 2 p +2 k-1)  }
\sum_{l=0}^k  \frac{ \Gamma ( 2 p+k+l -1)}{ \Gamma (k-l+1) \Gamma ( p+l )^2}
\frac{(-1)^{k+1}\Gamma(p-q+k)}{k!\Gamma(p-q)}
\\&=&
\sum_{l=0}^k  \frac{(-1)^l \Gamma (k+p)^2 \Gamma (k+l+2 p-1) \Gamma (l+p-q)}{l! \Gamma (k-l+1) \Gamma (2 k+2 p-1) \Gamma (l+p)^2 \Gamma (p-q)}~,
\label{summandCoo}
\eeqn
where   we set $ \Delta_0=p$. This   indeed equals to the coefficient $\sfC_{p,q}(k)$ in 
 \eqref{cpqn}, once we plug in the defining series   \eqref{hyperGfcn} of $_3F_2$ into  \eqref{cpqn}, which actually reduces to a finite sum, and compare each summand term by term with \eqref{summandCoo}. 
 
As a consequence, we obtain the following Mellin block expansion of celestial Mellin amplitude  \eqref{CAMeg}
 \be\label{Mlsbkexp}
 \cM(s)=\sum_{k=0}^\infty\frac{16^k \Gamma \left(k+\frac{1}{2}\right)^2}{\Gamma \left(\frac{1}{2}-k\right)^2 \Gamma (4 k+1)}
M_{2k+1}(s) 
 +\sum_{n=0}^\infty  \Big(1+(-1)^n e^{\pi i \alpha} \Big)\sfC_{1+\alpha, \frac12+\alpha} (n)\, M_{n+1+\alpha}(s) ~,\quad
 \ee
 which is consistent with conformal block expansion \eqref{cfblockexp}, and  indicates that   operators with $\Delta\in 2   \mathbb Z_{>0}$ are indeed absent.

 \subsection{Coefficient functions }

As we discussed before, the CFT data are equivalently encoded in the coefficient functions $I_\Delta, \tilde I_m$ which can be computed from inversion formulae.  Given the celestial correlator \eqref{cAmpeg}, we would like to perform the computation and find the explicit expressions of $I_\Delta, \tilde I_m$.  
\footnote{In {\cite{Garcia-Sepulveda:2022lga}}, the authors computed the  coefficient functions   of the imaginary part of a similar celestial amplitude. Here we will compute the coefficient functions $I_\Delta, \tilde I_m$ for the full celestial amplitude    \eqref{cAmpeg}.
}
We leave the technical details to Appendix~\ref{AppInversion}, and only quote the final results here. 

For the principal series, $I_\Delta$ is given by
\be\label{IDeltamodel}
I _\Delta =  \sfJ_\Delta (1, \frac12)
+ \sfJ _\Delta (1+\alpha, \frac12)+ \sfJ _\Delta (1+\alpha, \frac12+\alpha)~, \qquad
\sfJ_\Delta(p,q) =2\sfJ^0_\Delta(p,q)+\sfJ^+_\Delta(p,q) ~.
\ee
where the two functions $\sfJ_\Delta^0, \sfJ^+_\Delta$ take the following explicit forms:
  \beqn  
\sfJ_\Delta^+(p,q) &=&
 2 \pi  \csc (\pi  \Delta ) 
  B(p -\frac12-\alpha,-q+1)
  \Big[   \, _3F_2(p-\frac12-\alpha,1-\Delta ,\Delta ;1,p+\frac12-\alpha-q;1)
 \nonumber \\&&\qquad\qquad\qquad\qquad\qquad\qquad\qquad \quad
  + \, _3F_2( -q+1,1-\Delta ,\Delta ;1,p+\frac12-\alpha-q;1)\Big]~,\qquad\quad
 \\
\sfJ_\Delta^0(p,q) &=& \frac{(\sec (\pi  \Delta )+1) \Gamma (\Delta )^2 \Gamma (1-q) \Gamma (p+\Delta -1)}{\Gamma (2 \Delta ) \Gamma (p-q+\Delta )}
  {{}_3F_2 (\Delta, \Delta, -1+p+\Delta; 2\Delta,  p-q; 1)}{}
\nonumber  \\&&
  +\Big\{ \Delta \to 1-\Delta\Big\}~. 
\eeqn

For the discrete series, $\tilde I_m$ is given by
\beqn
\widetilde{I}_m &=& 
\frac{\pi  2^{3-2  m  } \Gamma ( m  )^2}{\Gamma \left( m  +\frac{1}{2}\right)^2}
\Bigg[ 
 \, _3F_2\left( m  , m  , m  ; m  +\frac{1}{2},2  m  ;1\right)
\nonumber\\& &  \qquad\qquad\qquad\quad +
\, _3F_2\left(\alpha + m  , m  , m  ;\alpha + m  +\frac{1}{2},2  m  ;1\right)
\frac{  \Gamma \left( m  +\frac{1}{2}\right) \Gamma (\alpha + m  ) }{\Gamma ( m  ) \Gamma \left(\alpha + m  +\frac{1}{2}\right)} e^{i \pi  \alpha }
\nonumber\\& &  \qquad\qquad\qquad\quad +
\, _3F_2\left(\alpha + m  , m  , m  ; m  +\frac{1}{2},2  m  ;1\right)
\frac{  \Gamma \left(\frac{1}{2}-\alpha \right) \Gamma (\alpha + m  )  }{\sqrt{\pi } \Gamma ( m  )} \big(\sin (\pi  \alpha )+1\big)
\Bigg]~.\qquad\qquad
 \eeqn

We see that the coefficient functions corresponding to   {\eqref{cAmpeg}} are quite complicated. In contrast, the celestial Mellin amplitude \eqref{CAMeg} takes a much simpler form.  

Despite  the complicated expressions of the coefficient functions, we still would like to understand some of the structures.  As discussed in  Appendix~\ref{AppInversion},  we find  the following residue formula \eqref{ResCpq}
\be
\Res_{\Delta=p+n}\frac{ \sfJ _\Delta}{2K_\Delta}=
-\sfC_{p,q}(n), \qquad n \in \mathbb N~.
\ee
 Together with \eqref{IDeltamodel}, this gives rise to the conformal block expansion which is the same as \eqref{cfblockexp}. 
 
  Furthermore, we have verified the following formula \eqref{IMpl}
 \beqn\label{IMpl}
\Res_{\Delta=m}\frac{I_\Delta }{K_\Delta}
=\frac{\Gamma (m)^2}{2 \pi ^2 \Gamma (2 m-1)}\tilde I_m~, \qquad m \in 2 \mathbb Z_{>0}~.
\eeqn
This  shows that there is a perfect cancellation between principal and discrete series at positive even integers. Using \eqref{OPEcoef},  we learn  that   operators with  positive even integer scaling dimensions should be absent  in exchange of $\phi$. This  is again in agreement with    \eqref{cOexchange}.

 \section{Comments on higher-dimensional generalizations} \label{CMAinhighD}

In this section, we will explore several possible definitions of celestial Mellin amplitude in four and higher dimensions. As we discussed in the introduction, the main technique difficulty comes from the delta function in celestial correlators due to momentum conservation. We will discuss how to overcome this subtlety. For simplicity, we will focus on the case of scalar fields in four dimensions.  
 
 Let us first review the  Mellin technique in standard CFTs. 
 In general, the Mellin amplitude for $n$-point scalar correlator in CFT is given by \cite{Mack:2009mi}
 \be
 \cG(x_i) =\EV{\cO_1(x_1) \cdots \cO_n(x_n)}
 =\int [d\gamma] \cM(\gamma_{ij}) \prod_{1\le i<j\le n} \Gamma(\gamma_{ij}) (x^2_{ij})^{-\gamma_{ij}}~,
 \ee
 where
 \be
 \sum_{j=1}^n \gamma_{ij}=0~, \qquad \gamma_{ii}=-\Delta_i~, \qquad \gamma_{ij} =\gamma_{ji}~.
 \ee
 
 Specializing to four   scalar operators with the same conformal dimensions $\Delta$, the correlation function is encoded in a function of cross-ratios:
  \be
 \EV{\cO (x_1) \cdots \cO (x_4)}=\frac{G(u,v)}{(x_{13}^2 x_{24}^2)^\Delta}~,
 \ee
 where the cross-ratios are given by
 \be
  u=
 z \bar z=\frac{x_{12}^2 x_{34}^2 }{x_{13}^2 x_{24}^2 }, \qquad
 v =
(1-z) (1-\bar z)=\frac{x_{14}^2 x_{23}^2 }{x_{13}^2 x_{24}^2 }~.
 \ee

 The Mellin amplitude is then defined by
  \be\label{MellinRep}
G(u,v) =\int_C \frac{dsdt}{(2\pi i)^2}
 u^{-s} v^{-t}
 \Gamma(s)^2 \Gamma(t)^2  \Gamma(\Delta-s -t)
\cM(s,t)~,
 \ee
 for  a properly chosen contour $C$. If the scalar operators are identical, we have the crossing symmetry $
 \cM(s,t )=\cM(t,s) =\cM(s, \Delta-s-t)
$.

 We would like to generalize our previous discussions on celestial Mellin amplitude to higher dimensions. For simplicity, let us consider the four scalar scattering in 4D spacetime. As in  the 3D case, the celestial amplitude is given by Mellin transform of the momentum space scattering amplitude~\eqref{CAmp}. It can be simplified as
   \be\label{GCAmp}
    \cG(  x_i,\bar x_i) = \EV{\cO (x_1,\bar x_1) \cdots \cO (x_4,\bar x_4)}=  K(x_i,\bar x_i) \times
    G(z,\bar z)~,
  \ee
  where
  \be
 K(x_i,\bar x_i)
 =  \prod_{i<j} |x_{ij}|^{ {\frac{ \Delta_T} {3}-\Delta_i-\Delta_j}{ }} 
 ~, \qquad \qquad 
   \Delta_T=\sum_{i=1}^4 \Delta_i~,
   \ee
  and
  \be\label{Gzzbar}
  G(z,\bar z)=  X(z ) \sfG(z )~,
  \ee 
  with
  \beqn\label{Xz}
   X(z )&=&2^{-\Delta_T+2} \Big|\frac{z}{\sqrt{1-z}}\Big|^{-\frac{ \Delta_T}{3}} \delta(z-\bar z) ~,
\\
 \sfG(z )&=&
\int_0^\infty d\omega\; \omega^{\Delta_T-4}A\Big( \epsilon_1 \epsilon_2 \omega^2, -\frac1z\epsilon_1 \epsilon_2 \omega^2\Big)~.
\eeqn
 The notable feature in the above four point correlator is that in \eqref{Xz} we have a delta function which sets $z=\bar z$. This comes from the momentum conservation which enforces the four scattering particles   lie on a plane, whose intersection with 
celestial sphere  restricts the position of  four operator  insertions. 

We would like to define the celestial Mellin amplitude for the celestial correlator \eqref{GCAmp}.  The delta function we emphasize above poses some challenges for a straightforward generalization. Nevertheless, there are several options and we would like to discuss them below. 

In the first approach, we just ignore the issue of delta function and just apply the Mellin 
techniques of standard CFT. So we can insert $G=X\sfG$ \eqref{Gzzbar} into \eqref{MellinRep}, and  try to find the corresponding Mellin amplitude $\cM(s, t)$.
In particular, we must reproduce the delta function from $\cM(s, t)$ using \eqref{MellinRep}. This is very subtle and not obvious at all, due to the very singular behavior of delta function. Nevertheless, we need to mention that in the study of conformal block expansion \cite{Atanasov:2021cje}, the authors indeed succeeded in reproducing   such a kind of delta function. 

In the second approach, we can remedy the singular delta function behavior by performing the light transformation \cite{Hu:2022syq} or shadow transformation \cite{Fan:2021isc}  on some operators in \eqref{GCAmp}.  The integral transformation would remove the delta function in \eqref{Xz}, \footnote{It is worth mentioning that we can also remove the unwanted delta function by breaking the translational invariance, and then apply \eqref{MellinRep} directly on the celestial amplitude.} and  the corresponding  four point function would be  very well behaved. Then we can use the  Mellin technique  \eqref{MellinRep} to study the light/shadow  transformed correlator. 
 
In the final approach,  we can restrict to the subspace  $z=\bar z$ of CCFT, which actually gives rise to a 1D defect CFT. \footnote{The defect is actually a bit trivial, because we are not really inserting any one dimensional defect into CCFT}  The correlator in the defect CFT is then given by   \eqref{Gzzbar} \eqref{Xz} but with delta function removed, possibly up to some extra functors involving $z$ which can be fixed by one dimensional $SL(2, \mathbb R)$ conformal symmetry.  Then we can define the corresponding celestial Mellin amplitude using \eqref{Ghatt} and \eqref{MellinTsf}, just like the   case  of 3D spacetime we discussed before.  

These are the most natural three possible definitions of celestial Mellin amplitude in higher dimensions.  The first approach is very subtle due to the delta function; the last  approach seems to be the simplest but the physical meaning in terms of 4D physics is not clear. They are not independent, and should be related in some way because they originate from the same celestial amplitude \eqref{GCAmp}. A detailed investigation on these problems is left to the future.

 \section{Conclusion} \label{conclusion}
 
 In this paper, we initiate the  study  of CCFT in Mellin representation. For technical reasons, we mainly focus on the  3D spacetime.  We introduce the notion of celestial Mellin amplitude, which is given by Mellin transforming the celestial amplitude in coordinate \eqref{MellinTsf}. As a result,  the celestial Mellin amplitude is the double Mellin transformations of momentum space scattering amplitude. The celestial Mellin  amplitude is useful in understanding the conformal block expansion in CCFT. More precisely, one can read off  possible exchanged operators from the position of poles, and the OPE coefficients from the residues at the poles via equation \eqref{OPEcoef}.  This formula is supposed to  be also  useful in the general study of 1D CFTs and the corresponding AdS$_2$. For illustration, we consider a simple model of scalar fields in 3D. We compute the celestial Mellin amplitude and its Mellin block expansion. We also compare it with the coefficient functions calculated from inversion formulae. 
 
 There are many interesting questions which remain to be further explored. One of the most important questions is to get a better understanding of celestial Mellin amplitudes  in four and higher dimensions. Although we provide  several possible definitions in higher dimensions, it remains to work out the technical details, in particular for the case with spinning particles.  After all, we would be finally interested in the graviton scattering in spacetime. Nevertheless, a good starting point  would be to consider the 4D analogue of the example considered in  section~\ref{CMAexample} and apply our various prescriptions in section~\ref{CMAinhighD}.

 Even for 3D spacetime,  there are various generalizations to be further explored.  One  straightforward direction is to consider the case with fermions or scalars with multiple species. The other more non-trivial question is to generalize the celestial Mellin amplitudes to higher points.

The Mellin representation is especially powerful in bootstrapping holographic CFTs and the dual  AdS quantum gravity.  Here we introduce the concept of celestial Mellin amplitude, aiming at generalizing the bootstrap philosophy to flat spacetime. As a starting point,   we may consider the following    bootstrap question even in 3D spacetime. From the EFT point of view, we can generalize  the simple model in section~\ref{CMAexample} by including all different higher derivative interactions, compute the scattering   amplitude and map it to the celestial basis. This   helps us establish a dictionary     between bulk  interaction  vertices and boundary CCFT data. We can then use various consistency condition to   bootstrap CCFT and thus the bulk couplings. We believe our celestial Mellin amplitude would be a powerful tool in such  a kind  bootstrap question. 

The crucial issue in the bootstrap   above is to find the consistency conditions of CCFT, which are supposed to inherit  from the bulk unitarity, locality, causality, etc. Translating  these universal bulk principles into CCFT is  a vital question. 
Ultimately, we    should be able  to bootstrap quantum gravity in flat spacetime from CCFT, just like the case of AdS/CFT.

 \acknowledgments
The work was   supported by the Royal Society under the grant, “Relations, Transformations, and Emergence in Quantum Field Theory” and the STFC under the grant “Amplitudes, Strings and Duality. No new data were generated or analysed during this study.

  \appendix 
 \section{A proof on the inverse of infinite dimensional matrices  } \label{MNinverse}
 
 In this appendix, we show that the two infinite dimensional matrices $\bm M, \bm N$ given in \eqref{Mmatrix} \eqref{Nmatrix} are indeed the inverse of each other. 
 
 By definition, we need to show that $\bm M \bm N=\bm N \bm M=1$. So we should consider the product of two matrices
 \beqn
\sum_{l=k_2}^{k_1} N_{k_1 l } M_{lk_2}
&=&
\frac{\Gamma ( \Delta_0+k_1 )^2 \Gamma ( 2 \Delta_0+2k_2)}{\Gamma ( \Delta_0+k_2 )^2\Gamma ( 2 \Delta_0+2k_1-1)} 
\sum_{l=k_2}^{k_1}
\frac{(-1)^{l-k_2} \;\Gamma (2  \Delta_0+k_1+l-1)}{(l-k_2)! \;\Gamma (k_2-l+1) \;\Gamma (2  \Delta_0+k_2+l)}
\qquad
\\&=&
\frac{\Gamma ( \Delta_0+k_1 )^2 \Gamma ( 2 \Delta_0+2k_2)}{\Gamma ( \Delta_0+k_2 )^2\Gamma ( 2 \Delta_0+2k_1-1)} \times 
\frac{   \delta_{k_1\, k_2}}{2 \Delta_0+2k_2-1}
= \delta_{k_1\, k_2}~,
 \eeqn
 where we used the  following identity which can be proved with the help of {\tt Mathematica}:
 \be
 \sum_{\ell=0}^k \frac{ (-1)^\ell }{ \ell ! (k-\ell )!}\frac{\Gamma(  S+\ell +k-1)}{\Gamma(  S+\ell)}=\frac{1}{(k+ S-1) \Gamma (1-k) \Gamma (k+1)}
 =\begin{cases}
 \frac{1}{ S-1} ~, & k=0~, \\
 0~, & k\in \mathbb Z_{>0}~.
 \end{cases}
 \ee

Similarly, we find that 
\be
\sum_{l=k_2}^{k_1} M_{k_1 l } N_{lk_2}
=\frac{\Gamma ( \Delta_0+k_1 )^2  }{\Gamma ( \Delta_0+k_2 )^2\Gamma ( 2 \Delta_0+2k_1-1)} \times 
 \Gamma(2\Delta_0+k_1+k_2 )\; \delta_{k_1\, k_2}   
= \delta_{k_1\, k_2}~,
 \ee
after using the identity
 \be
 \sum_{\ell=0}^k
\frac{(-1)^{k-\ell +1} (k-S-2 \ell +1) \Gamma (k+S-1) \Gamma (-k+S+\ell -1)}{\Gamma (\ell +1) \Gamma (k-\ell +1) \Gamma (S+\ell )}
 =\begin{cases}
 \Gamma( S-1) ~, & k=0~, \\
 0~, & k\in \mathbb Z_{>0}~.
 \end{cases}
 \ee

\section{More details on coefficient functions and inversion formulae}\label{AppInversion}

In  section~\ref{inversionForm}, we briefly review the inversion formulae which can be used to compute the coefficient functions. In this appendix,  we provide more technical details and apply the inversion formula to our celestial correlator \eqref{cAmpeg}.

Let us first review the conformal partial wave which takes different forms in different regions \cite{Mazac:2018qmi}: 
\beqn\label{Psi0pm}
\Psi _{\Delta}(z)  =
\begin{cases} 
\Psi^{(0)}_{\Delta}(z) ~,& z\in(0,1)~,
\\
 \Psi^{(-)}_{\Delta}(z) = \Psi^{(0)}_{\Delta}\left(\mbox{$\frac{z}{z-1}$}\right)~,\qquad\qquad\qquad\qquad\quad\;\;\,
 & z\in(-\infty,0)~,
\\
\Psi^{(+)}_{\Delta}(z) =
\frac{1}{2}\left[\Psi^{(0)}_{\Delta}(z+i\epsilon)+\Psi^{(0)}_{\Delta}(z-i\epsilon)\right]~,
 & z\in(1,\infty)~,
\end{cases}
\eeqn
where
\be\label{KDelta}
\Psi^{(0)}_{\Delta}(z)  =  K_{1-\Delta}G_{\Delta}(z) + K_{\Delta}G_{1-\Delta}(z)~, \qquad
K_{\Delta} =
\frac{\sqrt{\pi }\Gamma \left(\Delta -\frac{1}{2}\right) \Gamma \left(\frac{1-\Delta }{2}\right)^2}{\Gamma (1-\Delta ) \Gamma \left(\frac{\Delta }{2}\right)^2}\,.
\ee

Actually $\Psi^{(+)}_{\Delta}(z) $ also has various explicit forms  \cite{Lam:2017ofc,Mazac:2018qmi}
\beqn
\Psi^{(+)}_{\Delta}(z) 
&=&
\frac{2 \Gamma \left(\frac{1-\Delta }{2}\right) \Gamma \left(\frac{\Delta }{2}\right)  }{\sqrt{\pi }}
\, _2F_1\left(\frac{1-\Delta }{2},\frac{\Delta }{2};\frac{1}{2};\frac{(2-z)^2}{z^2}\right)
\\&=&
2 \pi  \csc (\pi  \Delta ) \left[\, _2F_1\left(1-\Delta ,\Delta ;1;\frac{1}{z}\right)+\, _2F_1\left(1-\Delta ,\Delta ;1;\frac{z-1}{z}\right)\right]~.
\eeqn

We would like to use inversion formulae to compute $I_\Delta, \tilde I_m$. For $I_\Delta$, we use the Euclidean inversion formula \eqref{eq:euclInv0}. We can split the integration into three regions, and use the relations  \eqref{Psi0pm}  and \eqref{CFTcrossing}-\eqref{CFTcrossing3} of $\Psi, G$ in these three  regions. As a result, we can    rewrite \eqref{eq:euclInv0} as an integral just  in one region $0<z<1$: 
 \beqn
I_\Delta &=& \int _{-\infty}^{\infty}dz\; z^{-2}\,\Psi_{\Delta}(z)\, {G}(z)
\nonumber\\&=& 
\int _{-\infty}^{0} dz \;z^{-2}\,\Psi_{\Delta}^0(\frac{z}{z-1})\, G^{(0)}(\frac{z}{z-1})
+\int _{0}^{1} dz \;z^{-2}\,\Psi_{\Delta}^0(z)\, G^{(0)}(z)
+\int _{1}^{+\infty} dz \;z^{-2}\,\Psi_{\Delta}^+(z)\,z^{2\Delta_\phi}   G  ^{(0)}(\frac 1 z) 
\nonumber\\&=& 
-\int _{1}^{0} dw w^{-2}\,\Psi_{\Delta}^0(w)\, G^{(0)}(w)
+\int _{0}^{1} dz \;z^{-2}\,\Psi_{\Delta}^0(z)\, G^{(0)}(z)
-\int _{1}^{0} dw\,\Psi_{\Delta}^+(1/w)\,w^{-2\Delta_\phi}   G  ^{(0)}(w) 
\nonumber\\&=& 
2 \int _{0}^{1} dz \;z^{-2}\, \, G^0(z) 
  \Psi_{\Delta}^{(0)}(z)  
  +
 \int _{0}^{1} dz \;z^{-2}\, \, G^{(0)}(z) 
 \Psi_{\Delta}^{(+)}(\frac 1 z)\,z^{ \frac12-\alpha} 
\nonumber\\&\equiv&  
2I_\Delta^0+ I_\Delta^+~,
\eeqn 
 where $\alpha=2\Delta_\phi-\frac32$. 
 
Before applying the above formula to \eqref{cAmpeg}, let us first consider some integrals:
\beqn
\sfJ_\Delta^0(p,q) &=&\int_0^1 dz\; z^{-2}\Psi^{(0)}_\Delta(z) \frac{z^p}{ (1-z)^q }
\\&=&  K_{1-\Delta} \int_0^1 dz\; G_\Delta(z)  {z^{p-2}}{ (1-z)^{-q} } 
+\Big\{ \Delta \to 1-\Delta\Big\}
\nonumber\\&=&
\frac{\sqrt{\pi }\Gamma \left( \frac{1}{2}-\Delta  \right) \Gamma \left(\frac{ \Delta }{2}\right)^2}{\Gamma (\Delta ) \Gamma \left(\frac{1-\Delta }{2}\right)^2}
\frac{\Gamma(1-q)  \Gamma(-1+p+\Delta) }{\Gamma( p-q+\Delta) }
 {{}_3F_2 (\Delta, \Delta, -1+p+\Delta; 2\Delta,  p-q; 1)}{}
\nonumber\\&&
+\Big\{ \Delta \to 1-\Delta\Big\}
\nonumber \\&=&
 \frac{(\sec (\pi  \Delta )+1) \Gamma (\Delta )^2 \Gamma (1-q) \Gamma (p+\Delta -1)}{\Gamma (2 \Delta ) \Gamma (p-q+\Delta )}
  {{}_3F_2 (\Delta, \Delta, -1+p+\Delta; 2\Delta,  p-q; 1)}{}
\nonumber \\&&
  +\frac{(\sec (\pi  \Delta )+1) \Gamma (1-\Delta )^2 \Gamma (1-q) \Gamma (p-\Delta )}{\Gamma (2-2 \Delta ) \Gamma (p-q+1-\Delta )}
  {{}_3F_2 (1-\Delta, 1-\Delta, p-\Delta; 2-2\Delta,  p-q; 1)}{}~.
  \nonumber\\ \label{Idelta0}
\eeqn

In the above evaluations, we use the following formula
 \be\label{GDeltaInt}
\int_0^1 dz\; G_\Delta(z) z^a (1-z)^b 
=\frac{\Gamma(1+b)  \Gamma(1+a+\Delta) }{\Gamma(2+a+b+\Delta) }
 {{}_3F_2 (\Delta, \Delta, 1+a+\Delta; 2\Delta, 2+a+b; 1)}{}~,
\ee
which is valid for $\Real b >-1, \Real (a+\Delta)>-1$. 
 
 Similarly, we can also consider
 \beqn
&&\int_0^1 dz\; z^{-2}\Psi^{(+)}_\Delta(1/z) \frac{z^p}{ (1-z)^q }
\\&=& 2 \pi  \csc (\pi  \Delta )\Big[ \int_0^1 dz\; \, _2F_1(1-\Delta ,\Delta ;1;z) {z^{p-2}}{ (1-z)^{-q} } +
  \int_0^1 dz\; \, _2F_1(1-\Delta ,\Delta ;1;1-z) {z^{p-2}}{ (1-z)^{-q} }  \Big]
  \nonumber
\\&=&
2 \pi  \csc (\pi  \Delta )\Big[ \int_0^1 dz\; \, _2F_1(1-\Delta ,\Delta ;1;z) {z^{p-2}}{ (1-z)^{-q} } +
  \int_0^1 dz\; \, _2F_1(1-\Delta ,\Delta ;1;z) {(1-z)z^{p-2}}{ z^{-q} }  \Big]
  \nonumber
 \\&=&
2 \pi  \csc (\pi  \Delta ) 
  B(p-1,-q+1)\Big[   \, _3F_2(p-1,1-\Delta ,\Delta ;1,p-q;1)
  + \, _3F_2( -q+1,1-\Delta ,\Delta ;1,p-q;1)\Big]~,
      \nonumber
\eeqn
where we use the formula
 \be\label{FInt}
 \int_0^1 dz\; z^a (1-z)^b\, _2F_1(1-\Delta ,\Delta ;1;z)
 =B(a+1,b+1) \, _3F_2(a+1,1-\Delta ,\Delta ;1,a+b+2;1)~, \qquad
 \Real a, b>-1~.
 \ee
 
 Therefore we have
  \beqn\label{Iplusdelta}
\sfJ_\Delta^+(p,q) &=& \int _{0}^{1} dz \;z^{-2}\,  \frac{z^p}{ (1-z)^q }
 \Psi_{\Delta}^{(+)}(1/z)\,z^{ \frac1 2-\alpha} 
 \\&=&
 2 \pi  \csc (\pi  \Delta ) 
  B(p -\frac12-\alpha,-q+1)
  \Big[   \, _3F_2(p-\frac12-\alpha,1-\Delta ,\Delta ;1,p+\frac12-\alpha-q;1)
 \nonumber \\&&\qquad\qquad\qquad\qquad\qquad\qquad\qquad\quad
  + \, _3F_2( -q+1,1-\Delta ,\Delta ;1,p+\frac12-\alpha-q;1)\Big]~. 
     \nonumber\\
     \label{JDeltaP}
\eeqn

With the help of two integrals in \eqref{Idelta0}\eqref{JDeltaP}, it is straightforward to compute the principal series coefficient function $I_\Delta$ for   \eqref{cAmpeg}. The full result is given in \eqref{IDeltamodel}.

Next, we would like to find the poles and residues of  $I_\Delta  /K_\Delta=2 I_\Delta^ 0/K_\Delta+I_\Delta ^+/K_\Delta$, which encodes the OPE data \eqref{OPEcoef}. For simplicity, we will assume that $p,q,\alpha$ are generic, in order to avoid the overlapping of poles.

  Let us first consider the simple poles at   $\Delta=p+n$ with $n \in \mathbb N$, which come from $\Gamma(p-\Delta) $  in  \eqref{Idelta0}.  The residues there are given by
  \beqn
  \Res_{\Delta=p+n}  \frac{I_\Delta^0}{K_\Delta}
  &=&
   \frac{(-1)^n q \Gamma (-q)  }{\Gamma (n+1) \Gamma (-n-q+1)}
   \, _3F_2\Big(-n-p+1,-n-p+1,-n;-n-q+1,-2 (n+p-1);1\Big)
  \nonumber \\  &=&
\frac{(-1)^{3 n+1} \Gamma (-2 (n+p-1)) \Gamma (n+p)^2 }{\Gamma (n+1) \Gamma (p)^2 \Gamma (-n-2 p+2)} \, _3F_2\Big(p-q,n+2 p-1,-n;p,p;1\Big)
\nonumber  \\&=&
  -\sfC_{p,q}(n)~,
 \eeqn
 where   $\sfC_{p,q}(n)$ is given in \eqref{cpqn}.
 In the second equality above,  we used the identity 
 \beqn
&&\, _3F_2(-n-p+1,-n-p+1,-n;-n-q+1,-2 n-2 p+2;1) \nonumber
\\&= & 
\frac{(-1)^{2 n} \Gamma (-2 (n+p-1)) \Gamma (n+p)^2 \Gamma (-n-q+1) }{\Gamma (p)^2 \Gamma (1-q) \Gamma (-n-2 p+2)} \, _3F_2(p-q,n+2 p-1,-n;p,p;1)~, \quad
\eeqn
which can be derived by combining \cite{HyperGeo,Garcia-Sepulveda:2022lga} \be\label{F32id1}
\, _3F_2(a,b,-n;e,f;1)=(-1)^n
\frac{  \Gamma (f) \Gamma (b-f+1)  }{\Gamma (f+n) \Gamma (b-f-n+1)}\, _3F_2(b,e-a,-n;e,b-f-n+1;1)~, \qquad
\ee
and
 \beqn
\, _3F_2(a,b,-n;e,f;1)&=&
\frac{(-1)^n \Gamma (e) \Gamma (f) \Gamma (a-f+1) \Gamma (b-f+1) }{\Gamma (e+n) \Gamma (f+n) \Gamma (a-f-n+1) \Gamma (b-f-n+1)}
\\& &\times
 \, _3F_2(-f-n+1,a+b-e-f-n+1,-n;a-f-n+1,b-f-n+1;1)~.  \nonumber
\eeqn

 For generic  $p,q,\alpha$, it is easy to convince oneself that this type of poles at at   $\Delta=p+n$ does not appear in $\sfJ_\Delta^+$ \eqref{Iplusdelta}.  As a result, we find 
  \be\label{ResCpq}
 \Res_{\Delta=p+n}  \frac{\sfJ_\Delta (p,q) }{2K_\Delta}=
 \Res_{\Delta=p+n}  \frac{2\sfJ^0 _\Delta (p,q)+\sfJ _\Delta^+(p,q)}{2K_\Delta}
=-\sfC_{p,q}(n)~.
 \ee
If we deform the contour in \eqref{eq:opeFromIs} to enclose such a pole,   we find the OPE coefficients there \eqref{OPEcoef}, in agreement with \eqref{usefulFormula}. Note that the minus sign is due to the orientation of the contour.

 From the formulae above, it appears that there may be also other poles at integers or half integers. Since we have reproduced \eqref{usefulFormula}, we expect that all the rest of poles should be cancelled. Showing this explicitly is tedious, so we will not try to do it here. Instead, we will try to reproduce the important results in  \eqref{cOexchange} where all  the positive even integral conformal dimensions are absent.  This means that there is a complete cancellation between discrete and principal series. We will now try to show this explicitly. 
 
 For this aim, we need to know the discrete series coefficient function $\tilde I_m$. To have a non-trivial and interesting check, we will use the Lorentz inversion formula \eqref{LorentzInv}. 
For this purpose, let us first introduce the vital quantity,  namely the double discontinuity,  which is defined as \cite{Caron-Huot:2017vep,Mazac:2018qmi}
\be\label{doubleDiscon}
\dDisc [G(z)]=  {G}(z)-\frac{ {G}^{\curvearrowleft}(z)+ {G}^{\text{\rotatebox[origin=c]{180}{\reflectbox{$\curvearrowleft$}}}}(z)}{2}
= G^{(0)}(z)-\frac{  G ^{(+)}(z+i\epsilon)+  G^{(+)}(z-i\epsilon)}{2}  ~,\qquad z\in (0,1)~,
\ee
where  $\cG^{(0)}(z) $ is defined on $z\in(0,1)$, analytically continuing to the upper and lower half complex plane; while $G^{(+)}(z)$ is given by \eqref{CFTcrossing2} in terms of $G^{(0)}(z)$, defined  on   $\mathbb C\backslash (-\infty, 1)$ after analytic continuation. 
The meaning of $G ^{(+)}(z\pm i\epsilon)$ is that we can start with $G^{(+)}$ defined on $(1, \infty)$ (using its relation \eqref{CFTcrossing2}  with $G^{(0)}$   defined on $(0,1)$), and then perform analytic continuation to the points $z\pm i\epsilon$ which sit above and below the branch cut   $z\in (0,1)\subset (-\infty, 1)$.

Following the definition, we have 
\be
\dDisc [z^a (1-z)^b]= z^a (1-z)^b-z^{2\Delta_\phi-a-b}(1-z)^b \cos(\pi  b)~.
\ee
 Therefore, for the celestial correlator in \eqref{cAmpeg}, the double discontinuity is given by 
 \be
\dDisc [ G(z)]=\frac{z}{\sqrt{1-z}} \Bigg[1+e^{\pi i \alpha} z^\alpha
+(1+\sin \pi \alpha) \Big(\frac{z}{1-z}\Big)^\alpha \Bigg]~.
\ee
 
 Applying the discrete series Lorentz inversion formula  \eqref{LorentzInv}, we find
\beqn
\widetilde{I}_m &=& \frac{4\Gamma(m)^2}{\Gamma(2m)}\int _{0}^{1} dz \;z^{-2}G_{m}(z)\dDisc\!\left[\mathcal{G}(z)\right] 
\\
 &=& \frac{4\Gamma(m)^2}{\Gamma(2m)}\int _{0}^{1} dz \;z^{-2} 
 \frac{z}{\sqrt{1-z}} \Bigg[1+e^{\pi i \alpha} z^\alpha
+(1+\sin \pi \alpha) \Big(\frac{z}{1-z}\Big)^\alpha \Bigg]G_{m}(z)
\\&=& 
\frac{\pi  2^{3-2  m  } \Gamma ( m  )^2}{\Gamma \left( m  +\frac{1}{2}\right)^2}
\Bigg[ 
 \, _3F_2\left( m  , m  , m  ; m  +\frac{1}{2},2  m  ;1\right)
\nonumber\\& &  \qquad\qquad\qquad\quad +
\, _3F_2\left(\alpha + m  , m  , m  ;\alpha + m  +\frac{1}{2},2  m  ;1\right)
\frac{  \Gamma \left( m  +\frac{1}{2}\right) \Gamma (\alpha + m  ) }{\Gamma ( m  ) \Gamma \left(\alpha + m  +\frac{1}{2}\right)} e^{i \pi  \alpha }
\nonumber\\& &  \qquad\qquad\qquad\quad +
\, _3F_2\left(\alpha + m  , m  , m  ; m  +\frac{1}{2},2  m  ;1\right)
\frac{  \Gamma \left(\frac{1}{2}-\alpha \right) \Gamma (\alpha + m  )  }{\sqrt{\pi } \Gamma ( m  )} (\sin (\pi  \alpha )+1)
\Bigg]~,\qquad\qquad
\label{Imtilde2}
\eeqn
where we use \eqref{GDeltaInt} to perform the integral.

We want to show that the OPE coefficients at positive even positive integers vanish. This is equivalent to showing 
 \beqn\label{IMpl}
\Res_{\Delta=m}\frac{I_\Delta }{K_\Delta}
=\frac{\Gamma (m)^2}{2 \pi ^2 \Gamma (2 m-1)}\tilde I_m~.
\eeqn
$I_\Delta$ gets contribution from $I_\Delta ^0$ and $I_\Delta ^+$. For     $I_\Delta^0/K_\Delta$, there are simple poles at  positive even integers because of the zeros of $K_\Delta$ \eqref{KDelta}. 
Using \eqref{Idelta0}, it is straightforward to calculate the residues of $I_\Delta^0/K_\Delta$ at positive even integers; the final result is very similar to \eqref{Imtilde2} except for an overall factor and the $\sin\pi \alpha $ term. More precisely, we find the difference
 \beqn 
\Res_{\Delta=m}\frac{I^+_\Delta }{K_\Delta}
&=&\Res_{\Delta=m}\frac{I _\Delta }{K_\Delta}
-\Res_{\Delta=m}\frac{2I^0_\Delta }{K_\Delta}
=\frac{\Gamma (m)^2}{2 \pi ^2 \Gamma (2 m-1)}\tilde I_m
 -\Res_{\Delta=m}\frac{2I^0_\Delta }{K_\Delta}
 \\
 &=&
 \, _3F_2\left(\Delta ,\Delta ,\alpha +\Delta ;2 \Delta ,\Delta +\frac{1}{2};1\right)
 \frac{2^{3-2 \Delta }  \Gamma \left(\frac{1}{2}-\alpha \right) \Gamma (\Delta )^3 \Gamma (\alpha +\Delta ) }{\pi ^{3/2} \Gamma \left(\Delta -\frac{1}{2}\right) \Gamma (2 \Delta ) \Gamma \left(\Delta +\frac{1}{2}\right)} \sin (\pi  \alpha )  ~. \qquad \qquad
\label{IMpl2}
\eeqn

To show \eqref{IMpl} is true, we need to compute the residues of $ {I^+_\Delta }/{K_\Delta}$ and prove the equation \eqref{IMpl2}. We will show below that this is indeed the case. 
For the celestial correlator  \eqref{cAmpeg}, we have 
\be
I^+_\Delta = \sfJ^+_\Delta (1+\alpha, \frac12)+\sfJ^+_\Delta (1, \frac12)
+\sfJ^+_\Delta (1+\alpha, \frac12+\alpha)~.
\ee
For the first term, it can be evaluated explicitly using \eqref{JDeltaP}
\be
\sfJ_\Delta^+(1+\alpha,\frac12)=\frac{2 \pi ^3 \csc (\pi  \Delta )}{\Gamma \left(1-\frac{\Delta }{2}\right)^2 \Gamma \left(\frac{\Delta +1}{2}\right)^2} ~,
\ee
which has zeros at positive even integer points. So $\sfJ_\Delta^+(1+\alpha,\frac12) /K_\Delta$ has no  poles at positive even integers.

The rest two terms together combine to
\beqn\label{Ideltaplussum}
\sfJ_\Delta^+(1,\frac12)+\sfJ_\Delta^+(1+\alpha,\frac12+\alpha) &=&
\frac{2 \pi ^{3/2} \Gamma \left(\frac{1}{2}-\alpha \right) \csc (\pi  \Delta ) }{\Gamma (1-\alpha )}
 \Bigg[\, _3F_2\left(\frac{1}{2},1-\Delta ,\Delta ;1,1-\alpha ;1\right)
\nonumber \\ & &\qquad\qquad\qquad\qquad\qquad\quad
 +\, _3F_2\left(\frac{1}{2}-\alpha ,1-\Delta ,\Delta ;1,1-\alpha ;1\right)\Bigg]~, \qquad\qquad
\eeqn
Using identity \eqref{F32id1}, one  can show that  the two terms in the square bracket cancel    for positive even integers $\Delta\in 2 \mathbb Z_{>0}$. So near the   positive  even integers, we   expect that the two terms in the square bracket of \eqref{Ideltaplussum} have the following expansion  
\be\label{A1def}
[\cdots]=0 + (\Delta-m) A_1 +\cdots, \qquad m \in2 \mathbb   Z_{>0}~,
\ee
On the other hand, the overall coefficient in \eqref{Ideltaplussum} can be expanded as
\be
\frac{2 \pi ^{3/2} \Gamma \left(\frac{1}{2}-\alpha \right) \csc (\pi  \Delta ) }{\Gamma (1-\alpha  )K_\Delta}
=
\frac{2^{3-2 \Delta } \Gamma \left(\frac{1}{2}-\alpha \right) \Gamma (\Delta )}{\pi  \Gamma (1-\alpha ) \Gamma \left(\Delta -\frac{1}{2}\right)}
\frac{1}{(\Delta-m)^2}
-\#\frac{1}{(\Delta-m) }
+\cdots~.
\ee
As a result,  we find the residues at positive even integers are given by
\be
\Res_{\Delta=m}\frac{I_\Delta^+}{K_\Delta}=
\frac{2^{3-2 \Delta } \Gamma \left(\frac{1}{2}-\alpha \right) \Gamma (\Delta )}{\pi  \Gamma (1-\alpha ) \Gamma \left(\Delta -\frac{1}{2}\right)}A_1~.
\ee
If  \eqref{IMpl} or equivalently \eqref{IMpl2}  holds, then  we should have the following relation 
\beqn
&&
\lim   _{\Delta\to m}
 \frac{\, _3F_2\left(\frac{1}{2},1-\Delta ,\Delta ;1,1-\alpha ;1\right)+\, _3F_2\left(\frac{1}{2}-\alpha ,1-\Delta ,\Delta ;1,1-\alpha ;1\right)}
 {  \Delta -m }
\\&=&
\frac{\partial}{\partial \Delta}\Bigg[ \, _3F_2\left(\frac{1}{2},1-\Delta ,\Delta ;1,1-\alpha ;1\right)+\, _3F_2\left(\frac{1}{2}-\alpha ,1-\Delta ,\Delta ;1,1-\alpha ;1\right)\Bigg]\Bigg|_{\Delta=m}
\\&=&
 \sin (\pi  \alpha ) 
\frac{  \Gamma (1-\alpha ) \Gamma (m )^2  \Gamma (\alpha +m )    }{\sqrt \pi \Gamma (2 m ) \Gamma \left(m +\frac{1}{2}\right)}
\, _3F_2\left(m ,m ,\alpha +m ;2 m ,m +\frac{1}{2};1\right)~,
\qquad m \in2 \mathbb Z_{>0}~, \qquad
\eeqn
where we use the definition of $A_1=\lim_{\Delta\to m} [\cdots]/(\Delta-m)$ in \eqref{A1def}. 
We don't have an analytic proof, but we have verified it numerically for certain range of $\alpha$. To conclude, we have thus argued that \eqref{IMpl} holds for the celestial correlator studied in this paper \eqref{cAmpeg}, implying a perfect cancellation between principal and discrete series at positive even integers. Therefore, the operators with positive even integer  scaling dimensions are absent.

 \bibliographystyle{JHEP} 
 
\bibliography{BMS.bib} 
  
\end{document}